\documentclass{article}
\usepackage{amssymb}
\usepackage{amsmath}
\usepackage{amsfonts}
\usepackage{geometry}
\usepackage{graphicx}
\usepackage{theorem}

\newtheorem{theorem}{Theorem}[section]

{\theorembodyfont{\normalfont}
\newtheorem{algorithm}[theorem]{Algorithm}

}

{\theorembodyfont{\normalfont}

}

\newtheorem{corollary}[theorem]{Corollary}
{\theorembodyfont{\normalfont}

}
\newtheorem{definition}[theorem]{Definition}
{\theorembodyfont{\normalfont}
\newtheorem{example}[theorem]{Example}

}

{\theorembodyfont{\normalfont}

}

\newtheorem{proposition}[theorem]{Proposition}
{\theorembodyfont{\normalfont}
\newtheorem{remark}[theorem]{Remark}

}

\begin{document}

\title{Note on Reversion, Rotation and Exponentiation in Dimensions Five and
Six}
\author{E. Herzig \\
Department of Mathematical Sciences\\
University of Texas at Dallas\\
Richardson, TX 75083 \and V. Ramakrishna \\
Department of Mathematical Sciences\\
University of Texas at Dallas\\
Richardson, TX 75083\\
Corresponding Author \and M.Dabkowski \\
Department of Mathematical Sciences\\
University of Texas at Dallas\\
Richardson, TX 75080}
\date{}
\maketitle

\begin{abstract}
The explicit matrix realizations of the reversion anti-automorphism and the
spin group depend on the set of matrices chosen to represent a basis of $1$%
-vectors for a given Clifford algebra. On the other hand, there are
iterative procedures to obtain bases of $1$-vectors for higher dimensional
Clifford algebras, starting from those for lower dimensional ones. For a
basis of $1$-vectors for ${\mbox C}l(0,5)$, obtained by applying such
procedures to a basis of $1$-vectors for ${\mbox C}l(3,0)$ consisting of the
Pauli matrices, we find that the matrix form of reversion involves neither $%
J_{4}$, nor $\widetilde{J}_{4}$, where $J_{2n}=\left( 
\begin{array}{cc}
0_{n} & I_{n} \\ 
-I_{n} & 0_{n}%
\end{array}%
\right) $ and $\widetilde{J}_{2n}=J_{2}\oplus J_{2}\oplus \ldots \oplus J_{2}
$. However, by making use of the relation between $4\times 4$ real matrices
and the quaternion tensor product ($\mathbb{H}\otimes \mathbb{H}$), the
matrix form of reversion for this basis of $1$-vectors is identified. The
corresponding version of the Lie algebra of the spin group, ${\mbox spin}(5)$%
, has useful matrix properties which are explored. Next, the form of
reversion for a basis of $1$-vectors for ${\mbox C}l(0,6)$ obtained
iteratively from ${\mbox C}l(0,0)$ is obtained. This is then applied to the
task of computing exponentials of $5\times 5$ and $6\times 6$ real
skew-symmetric matrices in closed form, by reducing this to the simpler task
of computing exponentials of certain $4\times 4$ matrices. For the latter
purpose closed form expressions for the minimal polynomials of these $%
4\times 4$ matrices are obtained, without having to compute their
eigenstructure. Finally a novel representation of $Sp(4)$ is provided which
may be of independent interest. Among the byproducts of this work are
natural interpretations for some members of an orthogonal basis for $M(4,%
\mathbb{R}
)$ provided by the isomorphism with $\mathbb{H}\otimes \mathbb{H}$, and a
first principles approach to the spin groups in dimensions five and six.
\end{abstract}

\section{Introduction}

The anti-automorphism reversion is central to the theory of Clifford
algebras. While it is unambiguously defined at the level of abstract
Clifford algebras, its explicit form as an involution of the matrix algebra,
to which the Clifford Algebra in question is isomorphic to, very much
depends on the specific basis of matrices for $1$-vectors chosen to make
concrete this isomorphism. Since there are canonical iterations supplying
bases of $1$-vectors for higher dimensional Clifford algebras, starting from
well known bases of $1$-vectors for lower dimensional ones (such as the
Pauli matrices for ${\mbox C}l\left( 3,\text{ }0\right) $), it is natural to
endow these bases with a privileged status. Hence finding the form of
reversion and Clifford conjugation with respect to these bases is
interesting. For Clifford conjugation it is known \cite{perti} that there is
(usually more than one) a choice of basis of $1$-vectors for ${\mbox C}%
l\left( 0,\text{ }n\right) $, with respect to which Clifford conjugation's
matrix form is given by Hermitian conjugation. However, no such easily
stated result is available for the \emph{matrix form of reversion} on ${%
\mbox C}l\left( 0,\text{ }n\right) $.

Explicit expressions for these two anti-automorphisms are important for a
variety of applications. For instance, if we can identify what reversion and
Clifford conjugation look like for ${\mbox Cl}\left( p,\text{ }q\right) $ as
matrix involutions for a given basis of $1$-vectors, then it becomes easy to
write what reversion and Clifford conjugation look like with respect to the
canonical basis of $1$-vectors for ${\mbox Cl}\left( p+1,\text{ }q+1\right) $
obtained from the said basis of $1$-vectors for ${\mbox Cl}\left( p,\text{ }%
q\right) $. A second application, motivating this work, is that explicit
matrix forms of these $2$ involutions are very much needed for the success
of a useful technique for computing the exponentials of elements of $%
\mathfrak{so}\left( n,\text{ }%
\mathbb{R}
\right) $ (the Lie algebra of $n\times n$ real, antisymmetric matrices). We
note that this Lie algebra and its Lie group arise in several applications
such as robotics, electrical and energy networks, photonic lattice filters,
communication satellites etc., \cite{goongi,chirikjian,fourporti,selig}

Computing the exponential of a matrix is arguably one of the central tasks
of applied mathematics. In general, this is quite a thankless job, \cite%
{dubious}. However, for matrices with additional structure certain
simplifications may be available. In particular, the theory of Clifford
Algebras and spin groups enables the reduction of finding $e^{X}$, with $%
X\in \mathfrak{so}\left( n,\text{ }%
\mathbb{R}
\right) $, to the computation of $e^{Y}$, where $Y$ is the associated
element in the Lie algebra of the corresponding spin group. Frequently this
means dealing with a matrix of smaller size. In particular, the minimal
polynomial of $Y$ is \underline{typically of lower degree} than that of $X$.
This connection, perhaps folklore, seems to have escaped the notice of a
variety of practitoners. Let us first illustrate this via the famous
Euler-Rodrigues formula for $\mathfrak{so}\left( 3,\text{ }%
\mathbb{R}
\right) $.

\begin{example}
\label{EulerRodrigues}Let $X=\left( 
\begin{array}{ccc}
0 & -c & b \\ 
c & 0 & -a \\ 
-b & a & 0%
\end{array}%
\right) $ be a $3\times 3$ antisymmetric real matrix.

As is well known, $X$ has a cubic minimal polynomial, viz., $X^{3}+\lambda
^{2}X=0$, with $\lambda ^{2}=a^{2}+b^{2}+c^{2}$. Hence $e^{X}=I+\frac{\sin
\lambda }{\lambda }X+\frac{1-\cos \lambda }{\lambda ^{2}}X^{2}$. This is the
famous \emph{Euler-Rodrigues formula}. We will now show that this formula
coincides with the following procedure$:$

\begin{description}
\item[Step 1] Identify $\mathfrak{su}\left( 2\right) $ with $P$, the purely
imaginary quaternions, and $SU\left( 2\right) $ with the unit quaternions

\item[Step 2] Let $\psi :P\rightarrow \mathfrak{so}\left( 3,\text{ }%
\mathbb{R}
\right) $ be the map obtained by linearizing the covering map $\Phi
:SU\left( 2\right) \rightarrow SO\left( 3,\text{ }%
\mathbb{R}
\right) $, where $\Phi $ is the matrix of the map, which sends sends $v\in P$
to $gvg^{-1}$, with $g$ a unit quaternion.

\item[Step 3] Find $\psi ^{-1}(X)$. This is $\frac{1}{2}(ai+bj+ck)$.

\item[Step 4] Compute the exponential of $\psi ^{-1}(X)$. This is the unit
quaternion $p=cos(\frac{\lambda }{2})1+\frac{\sin (\frac{\lambda }{2})}{%
\lambda }(ai+bj+ck)$, with $\lambda =\sqrt{a^{2}+b^{2}+c^{2}}$.

\item[Step 5] Compute the matrix of the map $x\in P\rightarrow px\bar{p}\in P
$, with respect to the basis $\{i,$ $j,$ $k\}$.
\end{description}

The matrix computed in \emph{Step 5} coincides with the matrix provided by
the Euler-Rodrigues formula, $e^{X}=I+\frac{\sin \lambda }{\lambda }X+\frac{%
1-\cos \lambda }{\lambda ^{2}}X^{2}$. For instance, the first column of the
matrix is \emph{Step 5} is found by computing $pi\bar{p}$ and rewriting this
element of $P$ as a vector in $%
\mathbb{R}
^{3}$. Computing $pi\bar{p}$ we find, it is 
\begin{equation*}
pi\bar{p}=\cos ^{2}(\frac{\lambda }{2})i+\frac{\cos (\frac{\lambda }{2})\sin
(\frac{\lambda }{2})}{\lambda })(2cj-2bk)+\frac{\sin ^{2}(\frac{\lambda }{2})%
}{\lambda ^{2}}(a^{2}i-b^{2}i-c^{2}i+2ack+2abj)
\end{equation*}%
This can be rewritten as 
\begin{equation*}
\cos ^{2}(\frac{\lambda }{2})i+\frac{\sin (\lambda )}{\lambda }(cj-bk)+\frac{%
\sin ^{2}(\frac{\lambda }{2})}{\lambda ^{2}}\left[
(a^{2}+b^{2}+c^{2})i-2(b^{2}+c^{2})i+2ack+2abj\right] 
\end{equation*}%
This simplifies to 
\begin{equation*}
i+\frac{\sin \lambda }{\lambda }(cj-bk)+\frac{1-\cos \lambda }{\lambda ^{2}}%
[-(b^{2}+c^{2})i+ack+abj]
\end{equation*}%
Rewritten as a vector in $%
\mathbb{R}
^{3}$ it is 
\begin{equation*}
\left( 
\begin{array}{c}
1+\frac{1-\cos \lambda }{\lambda ^{2}}[-(b^{2}+c^{2})] \\ 
c\frac{\sin \lambda }{\lambda }+ab\frac{1-\cos \lambda }{\lambda ^{2}} \\ 
-b\frac{\sin \lambda }{\lambda }+ac\frac{1-\cos \lambda }{\lambda ^{2}}%
\end{array}%
\right) 
\end{equation*}%
which is precisely the first column of \emph{Euler-Rodrigues formula} for $%
e^{X}$.

Strictly speaking, the above calculation is not what stems from considering $%
{\mbox C}l\left( 0,\text{ }3\right) $, since the latter is the double ring
of the quaternions. However, it is an easy exercise to show that doing all
calculations in ${\mbox C}l\left( 0,\text{ }3\right) $ amounts to the same
calculation outlined in the five step procedure above.

Though not of immense computational superiority in this simple instance, it
worth noting that the exponentiation of a $3\times 3$ matrix has been
reduced to the exponentiation of a $2\times 2$ matrix in $\mathfrak{su}%
\left( 2\right) $, the Lie algebra of $2\times 2$ traceless, anti-Hermitian
matrices (equivalently of a purely imaginary quaternion). Such matrices have
quadratic minimal polynomials, unlike $X$ which has a cubic minimal
polynomial.
\end{example}

This methodology extends in general. We will restrict ourselves to ${\mbox C}%
l\left( 0,\text{ }n\right) $ for simplicity. The method proceeds as follows:

\begin{algorithm}

\begin{description}
\item[Step 1] Identify a collection of matrices which serve as a basis of $1$
vectors for the Clifford Algebra ${\mbox C}l\left( 0,\text{ }n\right) $.

\item[Step 2] Identify the explicit form of Clifford conjugation ($\phi ^{cc}
$) and the grade (or so-called main) automorphism on ${\mbox C}l\left( 0,%
\text{ }n\right) $, with respect to this collection of matrices.
Equivalently identify the explicit form of Clifford conjugation and
reversion ($\phi ^{rev}$) with respect to this collection of matrices.

\item[Step 3] Steps 1 and 2 help in identifying both the spin group $%
Spin\left( n\right) $ and its Lie algebra $\mbox spin\left( n\right) $, as
sets of matrices, within the \underline{same matrix algebra}, that the
matrices in Step 1 live in. Hence, one finds an matrix form for the double
covering $\Phi _{n}:\mbox Spin\left( n\right) \rightarrow SO\left( n,\text{ }%
\mathbb{R}
\right) $. This is given typically as the matrix, with respect to the basis
of $1$-vectors in Step 1, of the linear map $H\rightarrow ZH\phi ^{cc}(Z)$,
with $H$ a matrix in the collection of $1$-vectors in \emph{Step 1} and $%
Z\in \mbox Spin\left( n\right) $. This enables one to express $\Phi _{n}(Z)$
as a matrix in $SO\left( n,\text{ }%
\mathbb{R}
\right) $.

\item[Step 4] Linearize $\Phi _{n}$ to obtain Lie algebra isomorphism $\Psi
_{n}:\mbox spin\left( n\right) \rightarrow \mathfrak{so}\left( n,\text{ }%
\mathbb{R}
\right) $. This reads as $W\rightarrow YW-WY$, with $W$ once again a $1$%
-vector and $Y\in \mbox spin\left( n\right) $. Once again this leads to a
matrix in $\mathfrak{so}\left( n,\text{ }%
\mathbb{R}
\right) $ which is $\Psi _{n}(Y)$.

\item[Step 5] Given $X\in \mathfrak{so}\left( n,\text{ }%
\mathbb{R}
\right) $ find $\Psi _{n}^{-1}(X)=Y\in \mbox spin\left( n\right) $.

\item[Step 6] Compute the matrix $e^{Y}$ and use Step 3 to find the matrix $%
\Phi _{n}(e^{Y})$. This matrix is $e^{X}$.
\end{description}
\end{algorithm}

\noindent The key steps for the success of this algorithm are really \emph{%
Steps 1, 2} and \emph{3}.\medskip 

In the literature, the identification of ${\mbox Spin}\left( n\right) $, is
usually achieved by using the isomorphism between ${\mbox C}l(0,n-1)$ and
the even vectors in ${\mbox C}l\left( 0,\text{ }n\right) $, \cite%
{pertii,portei}. In other words, ${\mbox Spin}\left( n\right) $, is
identified as a subset of ${\mbox C}l\left( 0,\text{ }n-1\right) $. However,
this does not enable the finding of the matrix form of reversion. Similarly,
to use Algorithm $1.2$ above, one needs the $1$-vectors, the $2$-vectors
(since they intervene in the Lie algebra of the spin group) and ${\mbox Spin}%
\left( n\right) $ to be identified as explicit subcollections of matrices
within the \underline{same} matrix algebra that ${\mbox C}l\left( 0,\text{ }%
n\right) $ is isomorphic to. Therefore, once a basis of $1$-vectors as a
specific collection of matrices has been found, one needs to find what forms
Clifford conjugation and reversion take with respect to this collection for
the successful realization of the applications above. Even if a realization
of $1$-vectors of ${\mbox C}l\left( 0,\text{ }n\right) $, as a subset of ${%
\mbox C}l\left( 0,\text{ }n-1\right) $, is specified, one still needs a
prescription of how both ${\mbox Spin}\left( n\right) $ and ${\mbox spin}%
\left( n\right) $ act on this set of $1$-vectors. Furthermore, the latter
action should be the linearization of the former action for applicability to
the problem of finding exponentials of matrices in $\mathfrak{so}\left( n,%
\text{ }%
\mathbb{R}
\right) $. See Remark \ref{complicated} below for more on this issue.

In this note, therefore, we prefer to do all calculations within ${\mbox C}%
l\left( 0,\text{ }n\right) $. One virtue of this is that it is a first
principles approach to the problem of identifying the spin group and thus
has some \emph{didactical} advantages also.

As mentioned above, there are iterative constructions enabling one to find a
basis of $1$-vectors for ${\mbox C}l\left( 0,\text{ }n\right) $, starting
from certain obvious bases of $1$-vectors for lower-dimensional Clifford
algebras (the iterative constructions, pertinent to this work, are
summarized in Sec $2.3$). Hence, it seems natural to use these for Step 1 of
the last algorithm. Thus, it is significant to be able to find the matrix
forms for reversion with respect to such a basis of $1$-vectors for ${\mbox C%
}l\left( 0,\text{ }n\right) $.

In particular, we found to our initial chagrin that for a basis of $1$%
-vectors for ${\mbox C}l\left( 0,\text{ }5\right) $, obtained from the Pauli
basis $\{\sigma _{j}\mid j=1,$ $2,$ $3\}$ for ${\mbox C}l\left( 3,\text{ }%
0\right) $, reversion is \underline{not} given by $X\rightarrow M^{-1}X^{T}M$
for $M=J_{4}$ or $M=\widetilde{J}_{4}$, as one might expect from the
circumstance that ${\mbox Spin}\left( 5\right) $ is isomorphic to $Sp\left(
4\right) $ (the group of $4\times 4$ matrices which are both unitary and
symplectic).

To circumvent this difficulty, we use the isomorphism between $\mathbb{H}%
\otimes \mathbb{H}$ and $M\left( 4,\text{ }%
\mathbb{R}
\right) $ to find a skew-symmetric and orthogonal $M$, for which reversion
is indeed described by $X\rightarrow M^{-1}X^{T}M$. Furthermore, this
isomorphism also enables us to find a conjugation between this $M$ and $J_{4}
$, and thus produce a basis of $1$-vectors of ${\mbox C}l\left( 0,\text{ }%
5\right) =M\left( 4,\text{ }%
\mathbb{C}
\right) $, with respect to which ${\mbox Spin}\left( 5\right) $ is indeed
the standard representation of $Sp\left( 4\right) $. It is emphasized,
however, that it is not obvious how to obtain this latter basis from first
principles, and hence the detour through $\mathbb{H}\otimes \mathbb{H}$ is
really useful, apart from being of independent interest. See, Remark (\ref%
{BlockStructureofWideHatsp}), for instance, for another illustration of this
utility.

It turns out that one obstacle to reversion not involving either $J_{4}$ nor 
$\widetilde{J}_{4}$ is the presence of either of these matrices themselves
in the basis of $1$-vectors for ${\mbox C}l\left( 0,\text{ }5\right) $. Not
having a tool such as the {$\mathbb{H}$}$\otimes ${$\mathbb{H}$} isomorphism
in higher dimensions, we work very carefully to arrive at a basis of $1$%
-vectors for ${\mbox C}l\left( 0,\text{ }6\right) $ which contains neither $%
J_{8}$ nor $\widetilde{J}_{8}$. For this we start with the sole possible
basis for ${\mbox C}l\left( 0,\text{ }0\right) $ and apply a judicious
combination of the iterative procedures in Sec $2.3$, to find a desirable
basis of $1$-vectors for ${\mbox C}l\left( 0,\text{ }6\right) $. This then
very naturally leads to $SU\left( 4\right) $ being the covering group in
dimension $6$.

\begin{remark}
\label{complicated}In \cite{portei} the derivation of $SU\left( 4\right) $
as the spin group in dimension $6$, is carried out in Pgs $80$, $151$ and $%
264-265$. As mentioned before, the Clifford algebra that \cite{portei} works
with for this purpose is actually ${\mbox C}l\left( 0,\text{ }5\right) $. In
particular, on Pgs $264-265$, an embedding of $%
\mathbb{R}
^{6}$, - the $1$-vectors for ${\mbox C}l\left( 0,\text{ }6\right) $, in ${%
\mbox C}l\left( 0,\text{ }5\right) =M\left( 4,\text{ }%
\mathbb{C}
\right) $ is used. Specifically, $%
\mathbb{R}
^{6}$ is identified with $\mathbb{%
\mathbb{C}
}^{3}$ and then $(z_{0},$ $z_{1},$ $z_{2})\in \mathbb{%
\mathbb{C}
}^{3}$ is identified with the following matrix in $M\left( 4,\text{ }%
\mathbb{C}
\right) $ 
\begin{equation*}
X\left( z_{0},\text{ }z_{1},\text{ }z_{2}\right) =\left( 
\begin{array}{cccc}
\bar{z}_{2} & 0 & z_{0} & \bar{z}_{1} \\ 
0 & \bar{z}_{2} & z_{1} & -\bar{z}_{0} \\ 
-\bar{z}_{0} & -\bar{z}_{1} & z_{2} & 0 \\ 
-z_{1} & z_{0} & 0 & z_{2}%
\end{array}%
\right) 
\end{equation*}%
But then the action of $\mbox spin(6)=\mathfrak{su}\left( 4\right) $ cannot
be the usual one, viz., $A\in \mathfrak{su}\left( 4\right) $ sending the one
vector $X\left( z_{0},\text{ }z_{1},\text{ }z_{2}\right) $ to the matrix $%
AX\left( z_{0},\text{ }z_{1},\text{ }z_{2}\right) -X\left( z_{0},\text{ }%
z_{1},\text{ }z_{2}\right) A$, since the latter is not of the form $X\left(
w_{0},\text{ }w_{1},\text{ }w_{2}\right) $ for some triple $\left( w_{0},%
\text{ }w_{1},\text{ }w_{2}\right) \in \mathbb{%
\mathbb{C}
}^{3}$. Indeed, the $(1,$ $2)$ entry of $AX(z_{0},$ $z_{1},$ $z_{2})-X\left(
z_{0},\text{ }z_{1},\text{ }z_{2}\right) A$ is non-zero typically.
Alternatively, note that the trace of the matrix $AX\left( z_{0},\text{ }%
z_{1},\text{ }z_{2}\right) -X\left( z_{0},\text{ }z_{1},\text{ }z_{2}\right)
A$ is zero for all $A\in \mathfrak{su}\left( 4\right) $ and for all $(z_{0},$
$z_{1},$ $z_{2})\in \mathbb{%
\mathbb{C}
}^{3}$. On the other hand the trace of $X\left( w_{0},\text{ }w_{1},\text{ }%
w_{2}\right) $ is $4{\mbox Re}\left( w_{2}\right) $.

It is emphasized that \cite{portei} does not make the claim in the above
paragraph, and the matrix $X\left( z_{0},\text{ }z_{1},\text{ }z_{2}\right) $
is used therein for an entirely different reason, viz., to avail of the fact
that every element of $\mbox Spin\left( n\right) $ can be factorized as a
product of an element in $S^{n-1}$ (the unit sphere in $%
\mathbb{R}
^{n}$) and an element in $\mbox Spin\left( n-1\right) $. The association of
the matrix $X\left( z_{0},\text{ }z_{1},\text{ }z_{2}\right) $ to the triple 
$(z_{0},$ $z_{1},$ $z_{2})$ is indeed elegant and the associated
factorization is quite useful. However, for the purposes of this note it is
necessary to proceed from first principles and work directly with ${\mbox C}%
l\left( 0,\text{ }6\right) =M\left( 8,\text{ }%
\mathbb{R}
\right) $. It seems that this is also didactically simpler for these
purposes.

There is also an unexpected benefit from working in ${\mbox C}l\left( 0,%
\text{ }6\right) $. Specifically, by starting with the obvious basis for ${%
\mbox C}l\left( 0,\text{ }1\right) $ and mimicking for ${\mbox C}l\left( 0,%
\text{ }5\right) $, the iterative constructions for ${\mbox C}l\left( 0,%
\text{ }6\right) $, alluded to above, we arrive at a basis of $1$-vectors
for ${\mbox C}l\left( 0,\text{ }5\right) $ which sheds some light on the
matrix $X\left( z_{0},\text{ }z_{1},\text{ }z_{2}\right) $ - see Remark (\ref%
{complicated2}). Further, by slightly modifying this construction we find a
natural interpretation of yet another member of the $\mathbb{H}\otimes 
\mathbb{H}$ basis for $M\left( 4,\text{ }%
\mathbb{R}
\right) $.\medskip 
\end{remark}

Thus, one byproduct of this note is useful interpretations for at least 3
elements of a basis of orthogonal matrices for $M\left( 4,\text{ }%
\mathbb{R}
\right) $, yielded by its isomorphism to $\mathbb{H}\otimes \mathbb{H}$ are
provided. More generally, our work can be seen as showing the utility of
Clifford Algebras for questions in algorithmic/computational linear algebra.
Thus this note is in the spirit of \cite{cliffminpolyi,haconi,nii,
ni,niii,expistruc,expisufour,noncompactportion,minpolyi}.

The other component of this work is an explicit characterization of minimal
polynomials of matrices in the Lie algebra of the spin groups of dimensions
5 and 6. These expressions are constructive and do not require any knowledge
of the eigenvalues/eigenvectors of these matrices. Once one has access to
these minimal polynomials computing the exponentials of matrices in these
Lie algebras is facile. One can either use recursions for the coefficients
of the exponential or use simple Lagrange interpolation (since the matrices
in question are all evidently diagonalizable and thus their minimal
polynomials have distinct roots). As mentioned before it is often the case
that the minimal polynomials of matrices in the Lie algebra of the spin
group is far lower than that of the corresponding element in $\mathfrak{so}%
\left( n,\text{ }%
\mathbb{R}
\right) $. Example \ref{striking} provides a striking illutsration of this
circumstance. Of course, a natural question that could be asked is whether
one could not directly compute exponentials of elements of ${\mbox spin}%
\left( n\right) $, without passing to a matrix algebra representation of
them, e.g, without using the fact that ${\mbox spin}(6)=\mathfrak{su}\left(
4\right) $, for instance. Computing exponentials of matrices by computing
exponentials directly within Clifford algebras has indeed been proposed in 
\cite{cliffminpolyi}. However, it has been our experience that it is only by
passing to the matrix representation that we are able to avail of certain
simplifications. For example, the fact that only certain types of
polynomials can arise as the minimal polynomials of matrices in $\mathfrak{su%
}\left( 4\right) $ is not evident from the fact that it is isomorphic to ${%
\mbox spin}(6)$. A full analysis of the advantages/disadvanatges of passing
to the matrix representation is beyond the scope of this paper, though it
certainly is an interesting question to investigate.

The balance of this note is organized as follows. In the next section basic
notation and preliminary facts are presented. Section $3$ derives the
explicit form of the reversion map for ${\mbox C}l\left( 0,\text{ }5\right) $
with respect to a basis of $1$ vectors obtained iteratively from the Pauli
matrices. As a byproduct the matrix forms of Clifford conjugation and
reversion on ${\mbox C}l(1,6)$ are derived. An algorithm is then presented,
which uses the derived form of reversion on ${\mbox C}l\left( 0,\text{ }%
5\right) $ to exponentiate in closed form a matrix in $\mathfrak{so}\left( 5,%
\text{ }%
\mathbb{R}
\right) $ by reducing this to the exponentiation of a $4\times 4$ matrix in
a Lie algebra, denoted $\widehat{sp}\left( 4\right) $. Section $4$ derives
explicit forms for minimal polynomials of matrices in $\widehat{sp}\left(
4\right) $, thereby providing a complete solution to the problem of
exponentiation of matrices in $\mathfrak{so}\left( 5,\text{ }%
\mathbb{R}
\right) $. The block structure of elements of $\widehat{sp}\left( 4\right) $
is shown to be amenable for calculation of the quantities intervening in the
expressions for these minimal polynomials. Section $5$ obtains the form of
reversion on ${\mbox C}l\left( 0,\text{ }6\right) $ with respect to a basis
of $1$-vectors obtained iteratively from the sole possible basis for ${\mbox
C}l\left( 0,\text{ }0\right) $. This is then applied to provide an algorithm
for exponentiating a matrix in $\mathfrak{so}\left( 6,\text{ }%
\mathbb{R}
\right) $ by reducing it to the corresponding problem in $\mathfrak{su}%
\left( 4\right) $. The next section then provides a complete list of closed
form expressions for minimal polynomials of matrices in $\mathfrak{su}\left(
4\right) $. The succeeding section revisits reversion on ${\mbox C}l\left( 0,%
\text{ }5\right) $ and sheds light on the matrix $X\left( z_{0},\text{ }%
z_{1},\text{ }z_{2}\right) $ in Remark \ref{complicated} and also finds an
interpretation for yet another element of the $\mathbb{H}\otimes \mathbb{H}$
basis. The final section offers conclusions. An appendix is devoted to a
representation of matrices in $Sp\left( 4\right) $ which may be of
independent interest.

\section{Notation and Preliminary Observations}

\subsection{Notation}

We use the following notation throughout

\begin{description}
\item[N1] $\mathbb{H}$ is the set of quaternions, while $\mathbb{P}$ is the
set of purely imaginary quaternions. Let $K$ be an associative algebra. Then 
$M(n,K)$ is just the set of $n\times n$ matrices with entries in $K$. For $K=%
\mathbb{C}
\mathbf{,}$ $\mathbb{H}$ we define $X^{\ast }$ as the matrix obtained by
performing entrywise complex (resp. quaternionic) conjugation first, and
then transposition. For $K=%
\mathbb{C}
$, $\bar{X}$ is the matrix obtained by performing entrywise complex
conjugation.

\item[N2] $J_{2n}=\left( 
\begin{array}{cc}
0_{n} & I_{n} \\ 
-I_{n} & 0_{n}%
\end{array}%
\right) $. Associated to $J_{2n}$ are $i)$ $Sp\left( 2n\right) =\{X\in
M\left( 2n,\text{ }%
\mathbb{C}
\right) \mid X^{\ast }X=I_{n},$ $J_{2n}^{-1}X^{T}J_{2n}=J_{2n}\}$. $Sp\left(
2n\right) $ is a Lie group; and $ii)$ $sp\left( 2n\right) =\{X\in M\left( 2n,%
\text{ }%
\mathbb{C}
\right) \mid X^{\ast }=-X,$ $X^{T}J_{2n}=-J_{2n}X\}$. $sp\left( 2n\right) $
is the Lie algebra of $Sp\left( 2n\right) $. Note many authors write $%
Sp\left( n\right) $ instead of our $Sp\left( 2n\right) $.

\item[N3] $\widetilde{J}_{2n}=J_{2}\oplus J_{2}\oplus \ldots \oplus J_{2}$.
Thus $\widetilde{J}_{2n}$ is the $n$-fold direct sum of $J_{2}$. $\widetilde{%
J}_{2n}$, is of course, explicitly permutation similar to $J_{2n}$, but it
is important for our purposes to maintain the distinction. Accordingly $%
\widetilde{Sp}\left( 2n\right) =\{X\in M\left( 2n,\text{ }%
\mathbb{C}
\right) \mid X^{\ast }X=I_{n},$ $\widetilde{J}_{2n}^{-1}X^{T}\widetilde{J}%
_{2n}=\widetilde{J}_{2n}\}$. $\widetilde{Sp}\left( 2n\right) $ is a Lie
group; and $ii)$ $\widetilde{sp}\left( 2n\right) =\{X\in M\left( 2n,\text{ }%
\mathbb{C}
\right) \mid X^{\ast }=-X,$ $X^{T}\widetilde{J}_{2n}=-\widetilde{J}_{2n}X\}$%
. $\widetilde{sp}\left( 2n\right) $ is the Lie algebra of $\widetilde{Sp}%
\left( 2n\right) $. Other variants of $J_{4}$ are of importance to this
paper, and they will be introduced later at appropriate points (see Remark %
\ref{The3Js} below).

\item[N4] The Pauli Matrices are 
\begin{equation*}
\sigma _{x}=\sigma _{1}=\left( 
\begin{array}{cc}
0 & 1 \\ 
1 & 0%
\end{array}%
\right) ;\text{ }\sigma _{y}=\sigma _{2}=\left( 
\begin{array}{cc}
0 & -i \\ 
i & 0%
\end{array}%
\right) ;\text{ }\sigma _{z}=\sigma _{3}=\left( 
\begin{array}{cc}
1 & 0 \\ 
0 & -1%
\end{array}%
\right) 
\end{equation*}

\item[N5] $SO(n,$ $%
\mathbb{R}
)$ stands for the $n\times n$ real orthogonal matrices with determinant one. 
$\mathfrak{so}\left( n,\text{ }%
\mathbb{R}
\right) $ is its Lie algebra - the set of $n\times n$ real antisymmetric
matrices.

\item[N6] $SU\left( n\right) $ is the Lie group of unitary matrices with
unit determinant, and $\mathfrak{su}\left( n\right) $ is its Lie algebra -
the set of anti-Hermitian matrices with zero trace.

\item[N7] The matrix $K_{2l}$ is 
\begin{equation*}
K_{2l}=\left( 
\begin{array}{cc}
0_{l} & I_{l} \\ 
I_{l} & 0_{l}%
\end{array}%
\right) 
\end{equation*}%
This matrix will be useful for succinctly expressing Clifford conjugation in
certain dimensions.

\item[N8] $A\otimes B$ stands for the Kronecker product of $A$ and $B$. $%
\left\Vert X\right\Vert _{F}$, for a matrix $X$, is $\sqrt{{\mbox Tr}%
(X^{\ast }X)}=\sqrt{\sum \sum_{i,j}\left\vert x_{ij}\right\vert ^{2}}$.
\end{description}

\subsection{Reversion and Clifford Conjugation}

We will not give formal definitions of notions from Clifford algebras. \cite%
{pertii,portei} are excellent texts wherein these definitions are to be
found. We will content ourselves with the following:

\begin{definition}
\label{basics}

\begin{description}
\item[I)] The reversion anti-automorphism on a Clifford algebra, $\phi ^{rev}
$, is the linear map defined by requiring that $i)$ $\phi ^{rev}(ab)=\phi
^{rev}(b)\phi ^{rev}(a)$; $ii)$ $\phi ^{rev}(v)=v$, for all $1$-vectors $v$;
and iii) $\phi ^{rev}(1)=1$. For brevity we will write $X^{rev}$ instead of $%
\phi ^{rev}(X)$.

\item[II)] The Clifford conjugation anti-automorphism on a Clifford algebra, 
$\phi ^{cc}$, is the linear map defined by a requiring that $i)$ $\phi
^{cc}(ab)=\phi ^{cc}(b)\phi ^{cc}(a)$; $ii)$ $\phi ^{cc}(v)=-v$, for all $1$%
-vectors $v$; and $iii)$ $\phi ^{cc}(1)=1$. For brevity $\phi ^{cc}(X)$ will
be written in the form $X^{cc}$.

\item[III)] The grade automorphism on a Clifford algebra, $\phi ^{gr}$ is $%
\phi ^{rev}\circ \phi ^{cc}$. As is well known it is also true that $\phi
^{gr}=\phi ^{cc}\circ \phi ^{rev}$. Once again we write $X^{gr}$ for $\phi
^{gr}(X)$.

\item[IV)] $\mbox Spin\left( n\right) $ is the collection of elements $x$ in 
${\mbox C}l\left( 0,\text{ }n\right) $ satisfying the following
requirements: $i)$ $x^{gr}=x$, i.e., $x$ is even; $ii)$ $xx^{cc}=1$; and $%
iii)$ For all $1$-vectors $v$ in ${\mbox C}l\left( 0,\text{ }n\right) $, $%
xvx^{cc}$ is also a $1$-vector. The last condition, in the presence of the
first two conditions, is known to be superfluous for $n\leq 5$, \cite%
{pertii,portei}.
\end{description}
\end{definition}

\subsection{Iterative Constructions in Clifford Algebras}

Here will outline $3$ iterative constructions of $1$-vectors for certain
Clifford Algebras, given a choice of one vectors for another Clifford
Algebra, \cite{pertii,portei}:

\begin{description}
\item[\textbf{IC1}] ${\mbox C}l\left( p+1,\text{ }q+1\right) $ as $M\left( 2,%
\text{ }{\mbox C}l\left( p,\text{ }q\right) \right) $, where $M(2,\mathfrak{A%
})$ stands for the set of $2\times 2$ matrices with entries in an
associative algebra $\mathfrak{A}$: Suppose $\{e_{1},\ldots
,e_{p},f_{1},\ldots ,f_{q}\}$ is a basis of $1$-vectors for ${\mbox C}%
l\left( p,\text{ }q\right) $. So, in particular, $e_{k}^{2}=+1,$ $k=1,\ldots
,p$ and $f_{l}^{2}=-1,$ $l=1,\ldots ,q$. Then a basis of $1$-vectors for ${%
\mbox C}l\left( p+1,\text{ }q+1\right) $ is given by the following
collection of elements in $M(2,{\mbox C}l\left( p,\text{ }q\right) )$: 
\begin{equation*}
\left( 
\begin{array}{cc}
e_{k} & 0 \\ 
0 & -e_{k}%
\end{array}%
\right) ,\text{ }k=1,\ldots ,p;\text{ }\left( 
\begin{array}{cc}
0 & 1 \\ 
1 & 0%
\end{array}%
\right) ;\text{ }\left( 
\begin{array}{cc}
f_{l} & 0 \\ 
0 & -f_{l}%
\end{array}%
\right) ,\text{ }l=1,\ldots ,q;\text{ }\left( 
\begin{array}{cc}
0 & 1 \\ 
-1 & 0%
\end{array}%
\right) 
\end{equation*}%
The $1$ and the $0$ in the matrices above are the identity and zero elements
of ${\mbox C}l\left( p,\text{ }q\right) $ respectively.

\item[\textbf{IC2}] From ${\mbox C}l\left( p,\text{ }q\right) $ to ${\mbox C}%
l\left( p-4,\text{ }q+4\right) $, for $p\geq 4$: Suppose $\{e_{1},\ldots
,e_{p},f_{1},\ldots ,f_{q}\}$ is a basis of $1$-vectors for ${\mbox C}%
l\left( p,\text{ }q\right) $. Let us label this basis as $\{g_{i}\mid
i=1,\ldots ,n\}$. Thus, $g_{i}=e_{i},$ $i=1,\ldots ,p$ and $g_{p+j}=f_{j},$ $%
j=1,\ldots ,q$. Then, to obtain a basis of $1$-vectors for ${\mbox C}l\left(
p-4,\text{ }q+4\right) $, we first compute 
\begin{equation*}
g=e_{1}e_{2}e_{3}e_{4}
\end{equation*}%
Then a basis $\{h_{i}\mid i=1,\ldots ,p+q\}$ of $1$-vectors for ${\mbox C}%
l\left( p-4,\text{ }q+4\right) $ is obtained by setting 
\begin{equation*}
h_{i}=g_{i}g,i=1,\ldots ,4;h_{i}=g_{i},i>4
\end{equation*}

\item[\textbf{IC3}] From ${\mbox C}l\left( p,\text{ }q\right) $ to ${\mbox C}%
l\left( q+1,\text{ }p-1\right) $ if $p\geq 1$. Suppose $\{e_{1},\ldots
,e_{p},f_{1},\ldots ,f_{q}\}$ is a basis of $1$-vectors for ${\mbox C}%
l\left( p,\text{ }q\right) $. Then a basis $\{\epsilon _{1},\ldots ,\epsilon
_{q+1},\mu _{1},\ldots ,\mu _{p-1}\}$ is obtained by defining 
\begin{equation*}
\epsilon _{1}=e_{1},\epsilon _{k+1}=f_{k}e_{1},\text{ }k=1,\ldots ,q
\end{equation*}%
and 
\begin{equation*}
\mu _{k}=e_{k+1}e_{1},\text{ }k=1,\ldots ,p-1
\end{equation*}%
In this last basis, the $\epsilon $'s square to $+1$, while the $\mu $'s
square to $-1$.
\end{description}

\begin{remark}
\label{Choiceofedoesnotmatter}In the last construction \textbf{IC3} above,
the special role played by $e_{1}$ could have been played by any one of the $%
e_{k},$ $k=1,\ldots ,p$. This would yield different sets of bases of $1$%
-vectors for ${\mbox C}l\left( q+1,\text{ }p-1\right) $, starting from a
basis of $1$-vectors for ${\mbox C}l\left( p,\text{ }q\right) $. We will
make use of this observation in Sec $8$.
\end{remark}

\begin{remark}
\label{ccandrevononemore}If Clifford conjugation and reversion have been
identified on ${\mbox C}l\left( p,\text{ }q\right) $ with respect to some
basis of $1$-vectors, then there are explicit expressions for Clifford
conjugation and reversion on ${\mbox C}l\left( q+1,\text{ }p-1\right) $ with
respect to the basis of $1$-vectors described in iterative construction 
\textbf{IC1} above. Specifically if $X=\left( 
\begin{array}{cc}
A & B \\ 
C & D%
\end{array}%
\right) $, then we have 
\begin{equation*}
X^{CC}=\left( 
\begin{array}{cc}
D^{rev} & -B^{rev} \\ 
-C^{rev} & A^{rev}%
\end{array}%
\right) 
\end{equation*}%
while reversion is 
\begin{equation*}
X^{rev}=\left( 
\begin{array}{cc}
D^{cc} & B^{cc} \\ 
C^{cc} & A^{cc}%
\end{array}%
\right) 
\end{equation*}%
This is immediate from the definitions of reversion and Clifford conjugation.

It is useful to observe that if elements of ${\mbox C}l\left( p,\text{ }%
q\right) $ have been identified with $l\times l$ matrices, then 
\begin{equation*}
X^{cc}=J_{2l}^{-1}[\left( 
\begin{array}{cc}
A^{rev} & B^{rev} \\ 
C^{rev} & D^{rev}%
\end{array}%
\right) ]^{BT}J_{2l}
\end{equation*}%
and that 
\begin{equation*}
X^{rev}=K_{2l}^{-1}[\left( 
\begin{array}{cc}
A^{cc} & B^{cc} \\ 
C^{cc} & D^{cc}%
\end{array}%
\right) ]^{BT}K_{2l}
\end{equation*}%
where $K_{2l}$ is the matrix at the end of Section $2.1$, and if $X=\left( 
\begin{array}{cc}
Y & Z \\ 
U & V%
\end{array}%
\right) $ is a $2\times 2$ block matrix, then $X^{BT}=\left( 
\begin{array}{cc}
Y & U \\ 
Z & V%
\end{array}%
\right) $
\end{remark}

\subsection{$\protect\theta _{%
\mathbb{C}
}$ and $\protect\theta _{\mathbb{H}}$ matrices:}

Some of the material here is to be found in \cite{hhorni}, for instance.

\begin{definition}
\label{complexmatrices}Given a matrix $M\in M\left( n,\text{ }%
\mathbb{C}
\right) $, define a matrix $\theta _{%
\mathbb{C}
}(M)\in M\left( 2n,\text{ }%
\mathbb{R}
\right) $ by first setting $\theta _{%
\mathbb{C}
}(z)=\left( 
\begin{array}{cc}
x & y \\ 
-y & x%
\end{array}%
\right) $ for a complex scalar $z=x+iy$. We then define $\theta _{%
\mathbb{C}
}(M)=(\theta _{%
\mathbb{C}
}(m_{ij}))$, i.e., $\theta _{%
\mathbb{C}
}(M)$ is a $n\times n$ block matrix, with the $(i,j)$th block equal to the $%
2\times 2$ real matrx $\theta _{%
\mathbb{C}
}(m_{ij})$.
\end{definition}

\begin{remark}
\label{Cproperties}Properties of $\theta _{%
\mathbb{C}
}$ Some useful useful properties of the map $\theta _{%
\mathbb{C}
}$ now follow:

\begin{description}
\item[i)] $\theta _{%
\mathbb{C}
}$ is an $%
\mathbb{R}
$-linear map.

\item[ii)] $\theta _{%
\mathbb{C}
}(MN)=\theta _{%
\mathbb{C}
}(M)\theta _{%
\mathbb{C}
}(N)$

\item[iii)] $\theta _{%
\mathbb{C}
}(M^{\ast })=[\theta _{%
\mathbb{C}
}(M)]^{T}$

\item[iv)] $\theta _{%
\mathbb{C}
}(I_{n})=I_{2n}$

\item[v)] A useful property is the following: $X\in M\left( 2n,\text{ }%
\mathbb{R}
\right) $ is in the image of $\theta _{%
\mathbb{C}
}$ iff $X^{T}=\widetilde{J}_{2n}^{-1}X^{T}\widetilde{J}_{2n}$.
\end{description}
\end{remark}

\begin{remark}
We call an $X\in {\mbox Im}(\theta _{%
\mathbb{C}
})$, a $\theta _{%
\mathbb{C}
}$ matrix. It is tempting, but confusing, to call such matrices complex
matrices. Similarly, if $X\in M\left( 2n,\text{ }%
\mathbb{R}
\right) $ satisfies $X^{T}=-\widetilde{J}_{2n}^{-1}X^{T}\widetilde{J}_{2n}$,
it will be called an anti - $\theta _{%
\mathbb{C}
}$ matrix. These are precisely the linear anti-holomorphic maps on $%
\mathbb{R}
^{2n}$. Note the map $X\rightarrow \widetilde{J}_{2n}^{-1}X\widetilde{J}_{2n}
$ is an involution on $M\left( 2n,\text{ }%
\mathbb{R}
\right) $. Its $+1$ eigenspace is precisely the space of $\theta _{%
\mathbb{C}
}$ matrices and its $-1$ eigenspace is the space of anti-$\theta _{%
\mathbb{C}
}$ matrices. Thus, from general properties of involutions, $M\left( 2n,\text{
}%
\mathbb{R}
\right) $ is a direct sum of these two subspaces.\vspace*{2mm}
\end{remark}

\noindent Next, to a matrix with quaternion entries will be associated a
complex matrix. First, if $q\in \mathbb{H}$ is a quaternion, it can be
written uniquely in the form $q=z+wj$, for some $z,$ $w\in 
\mathbb{C}
$. Note that $j\eta =\bar{\eta}j$, for any $\eta \in 
\mathbb{C}
$. With this at hand, the following construction associating complex
matrices to matrices with quaternionic entries (see \cite{hhorni} for
instance) is useful:

\begin{definition}
Let $X\in M(n,\mathbb{H})$. By writing each entry $x_{pq}$ of $X$ as%
\begin{equation*}
x_{pq}=z_{pq}+w_{pq}j,\text{ }z_{pq},w_{pq}\in 
\mathbb{C}
\end{equation*}%
we can write $X$ uniquely as $X=Z+Wj$ with $Z,$ $W\in M\left( n,\text{ }%
\mathbb{C}
\right) $. Associate to $X$ the following matrix $\theta _{\mathbb{H}}(X)\in
M\left( 2n,\text{ }%
\mathbb{C}
\right) $: 
\begin{equation*}
\theta _{\mathbb{H}}(X)=\left( 
\begin{array}{cc}
Z & W \\ 
-\bar{W} & \bar{Z}%
\end{array}%
\right) 
\end{equation*}
\end{definition}

\begin{remark}
\label{IntertwineConjugation}Viewing an $X\in M\left( n,\text{ }%
\mathbb{C}
\right) $ as an element of $M(n,\mathbb{H})$ it is immediate that $jX=\bar{X}%
j$, where $\bar{X}$ is entrywise complex conjugation of $X$.
\end{remark}

\noindent Next some useful properties of the map $\theta _{\mathbb{H}}:M(n,%
\mathbb{H})\rightarrow M\left( 2n,\text{ }%
\mathbb{C}
\right) $ are collected.

\begin{remark}
\label{Hproperties} Properties of $\theta _{\mathbb{H}}$:

\begin{description}
\item[i)] $\theta _{\mathbb{H}}$ is an $%
\mathbb{R}
$-linear map.

\item[ii)] $\theta _{\mathbb{H}}(XY)=\theta _{\mathbb{H}}(X)\theta _{\mathbb{%
H}}(Y)$

\item[iii)] $\theta _{\mathbb{H}}(X^{\ast })=[\theta _{\mathbb{H}}(X)]^{\ast
}$. Here the $\ast $ on the left is quaternionic Hermitian conjugation,
while that on the right is complex Hermitian conjugation.

\item[iv)] $\theta _{\mathbb{H}}(I_{n})=I_{2n}$

\item[v)] A less known property is the following: $\Lambda \in M\left( 2n,%
\text{ }%
\mathbb{C}
\right) $ is in the image of $\theta _{\mathbb{H}}$ iff $\Lambda ^{\ast
}=J_{2n}^{-1}X^{T}J_{2n}$.
\end{description}
\end{remark}

\begin{remark}
We call an $\Lambda \in {\mbox Im}(\theta _{\mathbb{H}})$, a $\theta _{%
\mathbb{H}}$ matrix. In \cite{hhorni} such matrices are called matrices of
the \underline{quaternion type}. But we eschew this nomenclature for the
same reason as for avoiding the terminology complex matrices. Similarly, if $%
\Lambda \in M\left( 2n,\text{ }%
\mathbb{C}
\right) $ satisfies $\Lambda ^{\ast }=-J_{2n}^{-1}X^{T}J_{2n}$, we say $%
\Lambda $ is an anti-$\theta _{\mathbb{H}}$ matrix. Note, that the map $%
\Lambda \rightarrow J_{2n}^{-1}\bar{\Lambda}J_{2n}$ is an involution. The $+1
$ eigenspace of this involution is precisely the subspace of $\theta _{%
\mathbb{H}}$ matrices, while the $-1$-eigenspace is the subspace of anti-$%
\theta _{\mathbb{H}}$ matrices, and hence $M\left( 2n,\text{ }%
\mathbb{C}
\right) $ is a direct sum of these two subspaces.
\end{remark}

\subsection{Minimal Polynomials and Exponential Formulae:}

The minimal polynomial of a matrix $X\in M\left( n,\text{ }%
\mathbb{C}
\right) $ is the unique monic polynomial, $m_{X}(x)$, of minimal degree
which annihilates $X$. Minimal polynomials can, just as any other
annihilating polynomial, be used to compute functions of $X$. One typical
mode to do so is to use the annihilating polynomial to establish recurrences
for higher powers of $X$, and in turn for any analytic function of $X$.
Naturally the recurrences are simpler on the eye, when the minimal
polynomial is used. An alternative method is to use such polynomials and
interpolation techniques for constructing functions of $X$, \cite{hhornii}.
This method is particularly useful when it is known in advance that $X$ is
diagonalizable (the only case of pertinence to this paper). In this case the
roots of the minimal polynomial are distinct and the venerable Lagrange
interpolation technique yields the desired function. We will confine
ourselves to giving explicit formulae for $e^{X}$ when $m_{X}$ is one of the
four following polynomials. Both the recurrence method and the interpolation
method lead to the same representation for $e^{X}$ as one may confirm.

\begin{theorem}
\label{ExpFromMinPolyList}Let $X\in M\left( n,\text{ }%
\mathbb{C}
\right) $ be non-zero. Then we have

\begin{description}
\item[I)] If $m_{X}(x)=x^{2}+\lambda ^{2}$, with $0\neq \lambda \in 
\mathbb{R}
$, then $e^{X}=\cos (\lambda )I+\frac{\sin (\lambda )}{\lambda }X$.

\item[II)] If If $m_{X}=x^{2}+2i\gamma x+\lambda ^{2}$, with $\gamma ,$ $%
\lambda \in 
\mathbb{R}
$, both non-zero, then $e^{X}=e^{-i\gamma }[(\cos (\sigma )+\frac{i\gamma }{%
\sigma }\sin (\sigma ))I+\frac{\sin (\sigma )}{\sigma }X]$, where $\sigma $
is the positive square root of $\lambda ^{2}+\gamma ^{2}$.

\item[III)] If $m_{X}=x^{3}+c^{2}x$, with $0\neq c\in 
\mathbb{R}
$, then $e^{X}=I+\frac{\sin c}{c}X+\frac{1-\cos c}{c^{2}}X^{2}$.

\item[IV)] If $m_{X}(x)=x^{4}+\theta ^{2}x^{2}+\lambda ^{2}$, with $\theta ,$
$\lambda \in 
\mathbb{R}
$, both non-zero, and satisfying $\theta ^{4}>4\lambda ^{2}$, then 
\begin{equation*}
e^{X}=\frac{1}{b^{2}-a^{2}}\{(\frac{b\sin a-a\sin b}{ab})X^{3}+(\cos a-\cos
b)X^{2}+(\frac{b^{3}\sin a-a^{3}\sin b}{ab})X+(b^{2}\cos a-a^{2}\cos b)I\}
\end{equation*}%
Here $a$ and $b$ are positive square roots of positive numbers $a^{2}$ and $%
b^{2}$, which in turn are defined to be the unique positive solutions to $%
a^{2}+b^{2}=\theta ^{2};$ $a^{2}b^{2}=\lambda ^{2}$.
\end{description}
\end{theorem}

\begin{remark}
It is possible that a matrix may be the sum of commuting summands, each of
which has a low degree minimal polynomial, even though the original matrix
has a high degree minimal polynomial. Thus, the exponential of such matrices
can be quite easily found. Some instances of this phenomenon are to be found
in \cite{expisufour}\textrm{.}
\end{remark}

\subsection{$\mathbb{H}\otimes \mathbb{H}$ and $M\left( 4,\text{ }%
\mathbb{R}
\right) $}

The algebra isomorphism between between $\mathbb{H}\otimes \mathbb{H}$ and $%
M\left( 4,\text{ }%
\mathbb{R}
\right) $ (also denoted by $gl(4,%
\mathbb{R}
)$) may be summarized as follows:

\begin{itemize}
\item Associate to each product tensor $p\otimes q\in \mathbb{H}\otimes 
\mathbb{H}$, the matrix, $M_{p\otimes q}$, of the map which sends $x\in 
\mathbb{H}$ to $px\bar{q}$, identifying $%
\mathbb{R}
^{4}$ with $\mathbb{H}$ via the basis $\left\{ 1,\text{ }i,\text{ }j,\text{ }%
k\right\} $. Here, $\bar{q}=q_{0}-q_{1}i-q_{2}j-q_{3}k$

\item Extend this to the full tensor product by linearity. This yields an
associative algebra isomorphism between $\mathbb{H}\otimes \mathbb{H}$ and $%
M\left( 4,\text{ }%
\mathbb{R}
\right) $. Furthermore, a basis for $gl(4,%
\mathbb{R}
)$ is provided by the sixteen matrices $M_{e_{x}\otimes e_{y}}$ as $e_{x},$ $%
e_{y}$ run through $1,$ $i,$ $j,$ $k$.

\item We define conjugation on $\mathbb{H}\otimes \mathbb{H}$ by setting $%
\bar{p\otimes q}=\bar{p}\otimes \bar{q}$ and then extending by linearity.
Conjugation in $\mathbb{H}\otimes \mathbb{H}$ corresponds to matrix
transposition, i.e., $M_{\bar{p}\otimes \bar{q}}=(M_{p\otimes q})^{T}$. A
consequence of this is that any matrix of the form $M_{1\otimes p}$ or $%
M_{q\otimes 1}$, with $p,$ $q\in \mathbb{P}$ is a real antisymmetric matrix.
Similarly, the most general special orthogonal matrix in $M\left( 4,\text{ }%
\mathbb{R}
\right) $ admits an expression of the form $M_{p\otimes q}$, with $p$ and $q$
both unit quaternions.
\end{itemize}

\begin{remark}
\label{ComplexSymmetric}$M\left( 4,\text{ }%
\mathbb{C}
\right) $: Since any complex matrix can be written as $Y+iZ$, with $Y,Z$ in $%
M\left( n,\text{ }%
\mathbb{R}
\right) $, it follows that matrices in $M\left( 4,\text{ }%
\mathbb{C}
\right) $ also possess quaternionic representations. In particular a complex
symmetric matrix can be written as $M_{p\otimes i+q\otimes j+r\otimes k}$,
with $p,$ $q,$ $r\in \mathbb{%
\mathbb{C}
}^{3}$. It should be clear from the context whether $i$ is a complex number
or a quaternion, in this regard. For instance $iM_{i\otimes j}$ [or just $%
i(i\otimes j)$] is the complex matrix equalling the complex numer $i$ times
the real matrix $M_{i\otimes j}$.
\end{remark}

\begin{remark}
\label{The3Js}Three matrices from this basis for $M\left( 4,\text{ }%
\mathbb{R}
\right) $ provided by $\mathbb{H}\otimes \mathbb{H}$ are important for us.
They are:

\begin{itemize}
\item $M_{1\otimes j}$ is precisely $J_{4}$.

\item The matrix $M_{1\otimes i}$, which we denote by $\widehat{J}_{4}$.

\item The matrix $M_{j\otimes 1}$, which we denote by $\breve{J}_{4}$.
\end{itemize}

Note that $\widetilde{J}_{4}$ is not part of this basis. It is, of course,
permutation similar to $J_{4}$. Each of these $3$ matrices above is both
antisymmetric and special orthogonal. As will be seen later the first two
are explicitly similar by a special orthogonal matrix. The third is similar
to the other two, but not by a special orthogonal similarity.
\end{remark}

\subsection{Other Matrix Theoretic Facts}

Throughout this note many important matrices are expressible as Kronecker
products $A\otimes B$ and so, the following properties of Kronecker products
will be freely used:

\begin{itemize}
\item $(A\otimes B)(C\otimes D) = AC \otimes BD$. $(A\otimes B)^{T} =
A^{T}\otimes B^{T}$.

\item If $A$ and $B$ are square then ${\mbox Tr}(A\otimes B)={\mbox Tr}(A){%
\mbox Tr}(B)$.\vspace*{2mm}
\end{itemize}

\noindent \emph{Schur's Determinantal Formulae}: We will use the following
special case of Schur's Determinantal Formulae, \cite{hhorni}: Suppose $%
X_{2n\times 2n}$ is 
\begin{equation*}
X=\left( 
\begin{array}{cc}
A & B \\ 
C & D%
\end{array}%
\right) 
\end{equation*}%
with $A,$ $B,$ $C,$ $D$ all $n\times n$. Then if $B$ is invertible, $\det
(X)=(-1)^{n^{2}}\det (B)\det (C-DB^{-1}A)$.

\section{Reversion and Rotation in Dimension Five}

First a basis of $1$-vectors for ${\mbox C}l\left( 0,\text{ }5\right) $ will
be constructed by starting with the Pauli basis for ${\mbox C}l\left( 3,%
\text{ }0\right) $ and applying the iterative constructions \textbf{IC1} and 
\textbf{IC2} of Section $2.3$.

Thus, let $\left\{ Z_{1}=\sigma _{x},\text{ }Z_{2}=\sigma _{y},\text{ }%
Z_{3}=\sigma _{z}\right\} $ be a basis of $1$-vectors for ${\mbox C}l\left(
3,\text{ }0\right) $. Applying \textbf{IC1} to this yields the following
basis for ${\mbox C}l\left( 4,\text{ }1\right) $: 
\begin{equation*}
\epsilon _{1}=\left( 
\begin{array}{cc}
\sigma _{x} & 0 \\ 
0 & -\sigma _{x}%
\end{array}%
\right) ;\text{ }\epsilon _{2}=\left( 
\begin{array}{cc}
\sigma _{y} & 0 \\ 
0 & -\sigma _{y}%
\end{array}%
\right) ;\text{ }\epsilon _{3}=\left( 
\begin{array}{cc}
\sigma _{z} & 0 \\ 
0 & -\sigma _{z}%
\end{array}%
\right) ;\text{ }\epsilon _{4}=\left( 
\begin{array}{cc}
0 & I_{2} \\ 
I_{2} & 0%
\end{array}%
\right) ;\text{ }e_{1}=\left( 
\begin{array}{cc}
0 & I_{2} \\ 
-I_{2} & 0%
\end{array}%
\right) 
\end{equation*}

Next let us apply \textbf{IC2} of Sec $2.3$ to this last basis to arrive at
a basis for ${\mbox C}l\left( 0,\text{ }5\right) $. To that end we first
need the product $\epsilon _{1}\epsilon _{2}\epsilon _{3}\epsilon _{4}$. A
quick calculation shows 
\begin{equation*}
\epsilon _{1}\epsilon _{2}\epsilon _{3}\epsilon _{4}=i\left( 
\begin{array}{cc}
0 & I_{2} \\ 
-I_{2} & 0%
\end{array}%
\right) =iJ_{4}
\end{equation*}

Then \textbf{IC2} says that a basis of $1$-vectors for ${\mbox C}l\left( 0,%
\text{ }5\right) $ is $\left\{ F_{i}\mid i=1,\ldots ,5\right\} $, as
given in Table 1. 

\begin{table}[tbp] \centering%
\begin{tabular}{lllll}
$F_{1}$ & $\mathbf{=}$ & $(\epsilon _{1}\epsilon _{2}\epsilon _{3}\epsilon
_{4})\epsilon _{1}i\left( 
\begin{array}{cc}
0 & -\sigma _{x} \\ 
-\sigma _{x} & 0%
\end{array}%
\right) $ & $=$ & $\sigma _{x}\otimes (-i\sigma _{x})$ \\ 
$F_{2}$ & $=$ & $(\epsilon _{1}\epsilon _{2}\epsilon _{3}\epsilon
_{4})\epsilon _{2}\left( 
\begin{array}{cc}
0 & -i\sigma _{y} \\ 
-i\sigma _{y} & 0%
\end{array}%
\right) $ & $=$ & $\sigma _{x}\otimes (-i\sigma _{y})$ \\ 
$F_{3}$ & $=$ & $(\epsilon _{1}\epsilon _{2}\epsilon _{3}\epsilon
_{4})\epsilon _{3}\left( 
\begin{array}{cc}
0 & -i\sigma _{z} \\ 
-i\sigma _{z} & 0%
\end{array}%
\right) $ & $=$ & $\sigma _{x}\otimes (-i\sigma _{z})$ \\ 
$F_{4}$ & $=$ & $(\epsilon _{1}\epsilon _{2}\epsilon _{3}\epsilon
_{4})\epsilon _{2}\left( 
\begin{array}{cc}
iI_{2} & 0 \\ 
0 & -iI_{2}%
\end{array}%
\right) $ & $=$ & $i\sigma _{z}\otimes I_{2}$ \\ 
$F_{5}$ & $=$ & $e_{1}$ & $=$ & $J_{4}$%
\end{tabular}%
\caption{$1$ - vectors for ${\mbox C}l(0,5)$ }\label{TableKey_2}%
\end{table}%

Note that the presence of $J_{4}$ in the basis is \emph{unavoidable}, by
construction, since the presence of $e_{1}=J_{4}$ in a basis of $1$-vectors
for ${\mbox C}l\left( 4,\text{ }1\right) $ and hence in that for ${\mbox C}%
l\left( 0,\text{ }5\right) $ is required by construction.

Inspired by the expected role of $J_{4}$, we now seek an expression for
reversion on ${\mbox C}l\left( 0,\text{ }5\right) $ of the form 
\begin{equation*}
\Phi ^{rev}(X)=M^{-1}X^{T}M
\end{equation*}%
where $M$ is a real orthogonal antisymmetric matrix. The unavoidable
presence of $J_{4}$ in the basis of $1$-vectors , immediately implies that $%
M\neq J_{4}$ and $M\neq \tilde{J}_{4}$. Indeed, for these two choices of $M$%
, we find that $M^{-1}F_{5}^{T}M=M^{-1}J_{4}^{T}M=-F_{5}\neq F_{5}$. So an
alternative choice for $M$ is needed. Given that we are working $4\times 4$
matrices, we are lead inexorably to the $\mathbb{H}\otimes \mathbb{H}$ basis
for $M\left( 4,\text{ }%
\mathbb{R}
\right) $.

Slight experimentation reveals that 
\begin{equation*}
M=M_{1\otimes i}=%
\begin{pmatrix}
0 & 1 & 0 & 0 \\ 
-1 & 0 & 0 & 0 \\ 
0 & 0 & 0 & -1 \\ 
0 & 0 & 1 & 0%
\end{pmatrix}%
\end{equation*}%
does the job, i.e., $M_{1\otimes i}^{-1}F_{i}^{T}M_{1\otimes i}=F_{i},$ for
all $i=1,\ldots ,5$.

It is useful to note that $M_{1\otimes i}$ also equals the following two
matrices:

\begin{description}
\item[i)] $M_{1\otimes i}=J_{2}\oplus (-J_{2})$. Since $J_{2}^{-1}=-J_{2}$,
this immediately reveals that $M_{1\otimes i}^{-1}=-M_{1\otimes i}$;

\item[ii)] $M_{1\otimes i}=\sigma _{z}\otimes (i\sigma _{y})$, and thus, $%
M_{1\otimes i}^{-1}=\sigma _{z}\otimes (-i\sigma _{y})$. This representation
is pertinent since the $F_{i}$ all have the form of Kronecker products of $%
2\times 2$ matrices and thus we will be able to use the properties of the
Kronecker product (see Section $2.7$) to facilitate calculation of $%
M_{1\otimes i}^{-1}F_{i}^{T}M_{1\otimes i}$.
\end{description}

The second of these two representations confirms that $\phi
^{rev}(X)=M_{1\otimes i}^{-1}X^{T}M_{1\otimes i}$. For future convenience we
denote $M_{1\otimes i}$ as $\hat{J}_{4}$, and correspondingly denote 
\begin{equation*}
\widehat{Sp}\left( 4\right) =\left\{ X\in M\left( 4,\text{ }%
\mathbb{C}
\right) \mid X\in U\left( 4\right) ,\ X^{T}\widehat{J}_{4}X=\widehat{J}%
_{4}\right\} 
\end{equation*}

It is well-known, and confirmed also by the above basis $\{F_{i}\}$, that
Clifford conjugation on ${\mbox C}l\left( 0,\text{ }5\right) $ is 
\begin{equation*}
\phi ^{cc}(X)=X^{\ast }
\end{equation*}

Hence the grade automorphism becomes 
\begin{equation*}
\phi^{gr}(X) = \widehat{J}_{4}^{-1}\bar{X}\widehat{J}_{4}
\end{equation*}

Thus, with respect to this choice of a basis of $1$-vectors, it is seen that 
\begin{equation*}
{\mbox Spin}\left( 5\right) =\left\{ X\in M\left( 4,\text{ }%
\mathbb{C}
\right) \mid X^{\ast }X=I_{4},\text{ }M_{1\otimes i}X=\bar{X}M_{1\otimes
i}\right\} =\widehat{Sp}\left( 4\right) 
\end{equation*}

In summary, we have shown the following:

\begin{proposition}
\label{Spin5IVersion} Let $B=\{F_{1}=\sigma _{x}\otimes (-i\sigma _{x}),$ $%
F_{2}=\sigma _{x}\otimes (-i\sigma _{y}),$ $F_{3}=\sigma _{x}\otimes
(-i\sigma _{z}),$ $F_{4}=i\sigma _{z}\otimes I_{2},$ $F_{5}=J_{4}\}$. Then $B
$ is a basis for $V$, the space of $1$-vectors for ${\mbox C}l\left( 0,\text{
}5\right) $. With respect to $B$ we have the following$:$

\begin{description}
\item[i)] The reversion anti-automorphism on ${\mbox C}l\left( 0,\text{ }%
5\right) $ is given by $\phi ^{rev}(X)=M_{1\otimes i}^{-1}X^{T}M_{1\otimes i}
$.

\item[ii)] Clifford conjugation is given by $\phi ^{cc}(X)=X^{\ast }$.

\item[iii)] $\mbox Spin\left( 5\right) =\widehat{Sp}\left( 4\right) =\left\{
X\in M\left( 4,\text{ }%
\mathbb{C}
\right) \mid X^{\ast }X=I_{4},\text{ }X^{T}\widehat{J}_{4}X=\widehat{J}%
_{4}\right\} $, where $\widehat{J}_{4}=M_{1\otimes i}$.

\item[iv)] The standard covering map $\Phi _{5}:\mbox Spin\left( 5\right)
\rightarrow SO\left( 5,\text{ }%
\mathbb{R}
\right) $ is given by sending $G\in \widehat{Sp}\left( 4\right) $ to the
matrix of the linear map $\Psi _{G}:V\rightarrow V$ where 
\begin{equation*}
\Phi _{G}(Y)=GYG^{\ast }
\end{equation*}%
with respect to the basis $B$.

\item[v)] The Lie algebra isomorphism $\Psi _{5}:\widehat{sp}\left( 4\right)
\rightarrow \mathfrak{so}\left( 5,\text{ }%
\mathbb{R}
\right) $, where $\widehat{sp}\left( 4\right) $ is the Lie algebra of the
group $\widehat{Sp}\left( 4\right) $, is obtained by linearizing $\Phi _{5}:%
\mbox Spin\left( 5\right) \rightarrow SO\left( 5,\text{ }%
\mathbb{R}
\right) $. Thus it is the map which sends $A\in \widehat{sp}\left( 4\right) $
to the matrix, with respect to $B$, of the linear map $\psi
_{A}:V\rightarrow V$ where 
\begin{equation*}
\psi _{A}(Z)=AZ-ZA
\end{equation*}
\end{description}
\end{proposition}

An immediate corollary of this result is that one can explicitly identify
the matrix forms of Clifford conjugation and reversion on ${\mbox C}l(1,6)$.

\begin{corollary}
\label{CCandrevonCliff16}Consider the following basis of $1$-vectors of ${%
\mbox C}l(1,6)=M(8,\mathbb{%
\mathbb{C}
})$, 
\begin{equation*}
\left\{ K_{8},\text{ }\left( 
\begin{array}{cc}
F_{i} & 0_{4} \\ 
0_{4} & -F_{i}%
\end{array}%
\right) ,\text{ }J_{8}\right\} 
\end{equation*}%
where $F_{i},$ $i=1,\ldots ,5$ is as in Proposition \ref{Spin5IVersion}.

Let $X\in M(8,\mathbb{%
\mathbb{C}
})={\mbox C}l(1,6)$. Then with respect to this basis of $1$-vectors we have

\begin{enumerate}
\item $X^{cc}=P^{-1}X^{T}P$, with $P=\left( 
\begin{array}{cc}
0_{4} & \hat{J}_{4} \\ 
-\hat{J}_{4} & 0_{4}%
\end{array}%
\right) $

\item $X^{rev}=K_{8}^{-1}X^{\ast }K_{8}$.
\end{enumerate}
\end{corollary}

\noindent \textbf{Proof:} This is an elementary consequence of block
multiplication and Remark \ref{ccandrevononemore}. $\diamondsuit $

\subsection{Computing the Lie Algebra Isomorphism $\protect\psi :\widehat{sp}%
\left( 4\right) \rightarrow \mathfrak{so}\left( 5,\text{ }%
\mathbb{R}
\right) $}

The Lie algebra of the $\widehat{Sp}\left( 4\right) $ is given by 
\begin{equation*}
\widehat{sp}\left( 4\right) =\left\{ X\in M\left( 4,\text{ }%
\mathbb{C}
\right) \mid X^{\ast }=-X,\text{ }X^{T}\hat{J_{4}}=-\hat{J_{4}}X\right\} 
\end{equation*}

The second condition is equivalent to saying that the $X\in \widehat{sp}%
\left( 4\right) $ can be expressed as $\widehat{J}_{4}S$, where $S$ is a 
\textit{complex} symmetric matrix. In view of Remark \ref{ComplexSymmetric},
this condition alone says that such an $X$'s $\mathbb{H}\otimes \mathbb{H}$
representation must be of the form 
\begin{equation*}
X=(1\otimes i)(p\otimes i+q\otimes j+r\otimes k+a1\otimes 1)
\end{equation*}%
with $p,$ $q,$ $r\in 
\mathbb{C}
^{3}$ and $a\in 
\mathbb{C}
$. However, the other condition, $X^{\ast }=-X$, forces $p\in 
\mathbb{R}
^{3},$ $a\in 
\mathbb{R}
$ and $q,$ $r\in (i%
\mathbb{R}
)^{3}$ (that is the components of $q,$ $r$ are purely imaginary).

Thus the most general such $X$ has an $\mathbb{H}\otimes \mathbb{H}$
representation of the form 
\begin{equation*}
X=-p\otimes 1+a1\otimes i+q\otimes k-r\otimes j
\end{equation*}%
with $p\in 
\mathbb{R}
^{3},$ $a\in 
\mathbb{R}
$ and $q,$ $r\in (i%
\mathbb{R}
)^{3}$. The negative signs are inessential and so a basis of $\widehat{sp}%
\left( 4\right) $ can be written in $\mathbb{H}\otimes \mathbb{H}$ form,
keeping in mind the remark on notation in Remark \ref{ComplexSymmetric}, as
in Table \ref{Table2}.

\begin{table}[tbp] \centering%
\begin{tabular}{lllllll}
$X_{1}$ & $=$ & $i(j\otimes j)$ &  & $X_{6}$ & $=$ & $i(i\otimes j)$ \\ 
$X_{2}$ & $=$ & $i\otimes 1$ &  & $X_{7}$ & $=$ & $1\otimes i$ \\ 
$X_{3}$ & $=$ & $k\otimes 1$ &  & $X_{8}$ & $=$ & $j\otimes 1$ \\ 
$X_{4}$ & $=$ & $i(j\otimes k)$ &  & $X_{9}$ & $=$ & $i(k\otimes k)$ \\ 
$X_{5}$ & $=$ & $i(k\otimes j)$ &  & $X_{10}$ & $=$ & $i(i\otimes k)$%
\end{tabular}%
\caption{Basis for $\widehat{sp}(4)$}\label{Table2}%
\end{table}%

Now to compute the image under $\Psi _{5}$ of such a basis element of $%
\widehat{sp}\left( 4\right) $, call it $X$, we have to compute $%
XF_{i}-F_{i}X,$ $i=1,\ldots ,5$ where $\{F_{i}\}$ is the basis of $1$%
-vectors in Proposition \ref{Spin5IVersion} and express the result as a real
linear combination of the $F_{i}$.

We will content ourselves with an illustration of the calculation for $%
X_{7}=1\otimes i$. We find

\begin{itemize}
\item $X_{7}F_{1}-F_{1}X_{7}=(\sigma _{z}\otimes i\sigma _{y})(\sigma
_{x}\otimes (-i\sigma _{x}))-(\sigma _{x}\otimes (-i\sigma _{x}))(\sigma
_{z}\otimes i\sigma _{y})=0$.

Here, the fact that $X_{7}$ can also be written as $(\sigma _{z}\otimes
i\sigma _{y})$ and that $F_{1}$ can also be written in the form $\sigma
_{x}\otimes (-i\sigma _{x})$ was employed.

\item $X_{7}F_{2}-F_{2}X_{7}=(\sigma _{z}\otimes i\sigma _{y})((\sigma
_{x}\otimes (-i\sigma _{y}))-(\sigma _{x}\otimes (-i\sigma _{y}))(\sigma
_{z}\otimes i\sigma _{y})=2\sigma _{z}\sigma _{x}\otimes I_{2}=2i\sigma
_{y}\otimes I_{2}=2F_{5}$.

\item $X_{7}F_{3}-F_{3}X_{7}=(\sigma _{z}\otimes i\sigma _{y})(\sigma
_{x}\otimes (-i\sigma _{z}))-(\sigma _{x}\otimes (-i\sigma _{z}))(\sigma
_{z}\otimes i\sigma _{y})=0$.

\item $X_{7}F_{4}-F_{4}X_{7}=(\sigma _{z}\otimes i\sigma _{y})(i\sigma
_{z}\otimes I_{2}-(i\sigma _{z}\otimes I_{2}(\sigma _{z}\otimes i\sigma
_{y})=0$.

\item $X_{7}F_{5}-F_{5}X_{7}=(\sigma _{z}\otimes i\sigma _{y})i\sigma
_{y}\otimes I_{2}-i\sigma _{y}\otimes I_{2}(\sigma _{z}\otimes i\sigma
_{y})=2\sigma _{x}\otimes (i\sigma _{y})=-2F_{5}$.
\end{itemize}

Hence $\Psi _{5}(X_{7})=\left( 
\begin{array}{ccccc}
0 & 0 & 0 & 0 & 0 \\ 
0 & 0 & 0 & 0 & -2 \\ 
0 & 0 & 0 & 0 & 0 \\ 
0 & 0 & 0 & 0 & 0 \\ 
0 & 2 & 0 & 0 & 0%
\end{array}%
\right) $. More compactly,%
\begin{equation*}
\Psi _{5}(X_{7})=2(e_{5}e_{2}^{T}-e_{2}e_{5}^{T})\text{ (here, ofcourse }%
e_{i}\text{ is the }i\text{th standard unit vector)}
\end{equation*}

In summary, the following holds:

\begin{theorem}
\label{TableforSoFiveLieIso}The Lie algebra isomorphism $\Psi _{5}:\widehat{%
sp}\left( 4\right) \rightarrow \mathfrak{so}\left( 5,\text{ }%
\mathbb{R}
\right) $ is described by Table \ref{TableKey_1}$:$

\begin{table}[tbp] \centering%
\begin{tabular}{lllll}
$\widehat{sp}\left( 4\right) $ & $\mathfrak{so}\left( 5,\text{ }%
\mathbb{R}
\right) $ &  & $\widehat{sp}\left( 4\right) $ & $\mathfrak{so}\left( 5,\text{
}%
\mathbb{R}
\right) $ \\ 
$iM_{j\otimes j}$ & $2(e_{1}e_{2}^{T}-e_{2}e_{1}^{T})$ &  & $iM_{i\otimes j}$
& $2(e_{2}e_{4}^{t}-e_{4}e_{2}^{T})$ \\ 
$M_{i\otimes 1}$ & $2(e_{3}e_{1}^{T}-e_{1}e_{3}^{T})$ &  & $M_{1\otimes i}$
& $2(e_{5}e_{2}^{T}-e_{2}e_{5}^{T})$ \\ 
$M_{k\otimes 1}$ & $2(e_{1}e_{4}^{T}-e_{4}e_{1}^{T})$ &  & $M_{j\otimes 1}$
& $2(e_{4}e_{3}^{T}-e_{3}e_{4}^{T})$ \\ 
$iM_{j\otimes k}$ & $2(e_{1}e_{5}^{T}-e_{5}e_{1}^{T})$ &  & $iM_{k\otimes k}$
& $2(e_{5}e_{3}^{T}-e_{3}e_{5}^{T})$ \\ 
$iM_{k\otimes j}$ & $2(e_{2}e_{3}^{T}-e_{3}e_{2}^{T})$ &  & $iM_{i\otimes k}$
& $2(e_{5}e_{4}^{T}-e_{4}e_{5}^{T})$%
\end{tabular}%
\caption{Lie algebra isomorphism between $\widehat{sp}(4)$ and
$\mathfrak{so}\left( 5,\text{ }\mathbb{R}\right) $ }\label{TableKey_1}%
\end{table}%
\end{theorem}

\begin{remark}
\label{ConjugacyHat}We have $\widehat{J}_{4}=M_{1\otimes i}$, while the
standard representation of the symplectic form, $J_{4}$ is $%
J_{4}=M_{1\otimes j}$. This makes it extremely easy to find a special
orthogonal conjugation between the two. Since every element of $SO(4,%
\mathbb{R}
)$ has a $\mathbb{H}\otimes \mathbb{H}$ representation of the form $%
M_{p\otimes q}$, for unit quaternions, we let $U^{T}=M_{p\otimes q}$ and
seek $U$ so that 
\begin{equation*}
U^{T}\widehat{J}_{4}U=J_{4}
\end{equation*}%
Using the properties of the isomorphism $\mathbb{H}\otimes \mathbb{H}\simeq
M\left( 4,\text{ }%
\mathbb{R}
\right) $ of Section $2.6$, it is obvious that we can let $p=1$ and seek $q$
to be a unit quaternion satisfying 
\begin{equation*}
qi\bar{q}=j
\end{equation*}

Of the infinite choices possible, let us pick $q=\frac{1}{\sqrt{2}}(1+k)$
for concreteness. The corresponding $U^{T}$ can then also be expressed as $%
\frac{1}{\sqrt{2}}(I_{4}+\sigma _{x}\otimes (i\sigma _{y}))$.

With this explicit conjugation available, the following are immediate:

\begin{description}
\item[I)] $U[Sp\left( 4\right) ]U^{T}=\widehat{Sp}\left( 4\right) $; and $%
U[sp\left( 4\right) ]U^{T}=\widehat{sp}\left( 4\right) $.

\item[II)] One can use this conjugation to find yet another basis of $1$
-vectors for ${\mbox C}l\left( 0,\text{ }5\right) $, viz., 
\begin{equation*}
\left\{ I_{2}\otimes (i\sigma _{z}),\text{ }\sigma _{x}\otimes (i\sigma
_{y}),\text{ }I_{2}\otimes (i\sigma _{x}),\text{ }i\sigma _{y}\otimes \sigma
_{y},\text{ }\sigma _{z}\otimes (i\sigma _{y})\right\} 
\end{equation*}%
With respect to this basis Clifford conjugation is once again Hermitian
conjugation, but reversion is $Y\rightarrow J_{4}^{-1}Y^{T}J_{4}$. Thus, $%
\mbox Spin\left( 5\right) $ is, with respect to this basis, the standard
representation of $Sp\left( 4\right) $.

\item We emphasize however, that this basis was arrived at only by going
through $\widehat{J}_{4}$ first. In other words, this basis, to the best of
our knowledge, does \underline{not} naturally arise from first principles as
does the basis $\left\{ F_{i}\mid i=1,\ldots ,5\right\} $ in Proposition \ref%
{Spin5IVersion}\textrm{. }
\end{description}
\end{remark}

\noindent \textbf{Computing Exponentials in $\mathfrak{so}(5,$ }$%
\mathbb{R}
$\textbf{$)$}

Specializing Algorithm $1.2$ yields the following method for computing the
exponential of a matrix in $\mathfrak{so}\left( 5,\text{ }%
\mathbb{R}
\right) $:

\begin{itemize}
\item If $X\in \mathfrak{so}\left( 5,\text{ }%
\mathbb{R}
\right) $, find $Y=\Psi _{5}^{-1}(X)\in \widehat{sp}\left( 4\right) $ using
Table 3.

\item Compute $e^{Y}$.

\item Find $e^{Y}F_{j}e^{-Y},$ $\forall j=1,\ldots ,5$. Express $%
e^{Y}F_{j}e^{-Y}=\sum_{i=1}^{5}c_{ij}F_{i}$.

\item Then $e^{X}$ is the matrix whose $i$th column is $\left (%
\begin{array}{c}
c_{i1} \\ 
c_{i2} \\ 
\vdots \\ 
c_{i5}%
\end{array}
\right )$.
\end{itemize}

Thus, the problem of computing $e^{X}$ is reduced to the problem of
computing the exponential of a $4\times 4$ matrix, $Y$, which furthermore
has additional structure, thereby rendering the computation of $e^{Y}$ in
closed form very easy.

\section{Minimal Polynomials of Matrices in $\widehat{sp}\left( 4\right) $}

In this section we show that the minimal polynomials of matrices in $Y\in 
\widehat{sp}\left( 4\right) $ can be computed explicitly, and that these
explicit forms lead correspondingly to explicit formulae for $e^{Y}$.
Indeed, as will be seen below, the minimal polynomials that arise are each
one of the four types in Theorem \ref{ExpFromMinPolyList}.

To this end, it is easier to work with matrices in the standard
representation, viz., $sp\left( 4\right) $, and use the connection of such
matrices to $M\left( 2,\text{ }\mathbb{H}\right) $. It should be pointed
that the results obtained below are invariant under conjugation by a special
orthogonal matrix, and hence extend verbatim to matrices in $\widehat{sp}%
\left( 4\right) $ and thus there is no need to find first
the element in $sp\left(
4\right) $ conjugate to the matrix $Y\in \widehat{sp}\left( 4\right) $ (See
Remark \ref{NoNeedtoConjugateY}). In fact, it will be seen in Remark \ref%
{BlockStructureofWideHatsp} that the quantities intervening in the result
about the minimal polynomials are easier to calculate for $\widehat{sp}%
\left( 4\right) $.

Recall that if $Z\in M\left( 2,\text{ }\mathbb{H}\right) $, then $Z=A+Bj$,
with $A,$ $B\in M\left( 2,\text{ }%
\mathbb{C}
\right) $. Denote 
\begin{equation*}
Y=\theta _{\mathbb{H}}(Z)=\left( 
\begin{array}{cc}
A & B \\ 
-\bar{B} & \bar{A}%
\end{array}%
\right) 
\end{equation*}%
Hence by $v)$ of Remark \ref{Hproperties} of Sec $2.4$, 
\begin{equation*}
Y^{\ast }=Y^{\dagger }
\end{equation*}%
where $Y^{\dagger }=-J_{4}Y^{T}J_{4}$. Matrices in $sp\left( 4\right) $ are
clearly $\theta _{\mathbb{H}}$-matrices. Therefore, the following result is
pertinent:

\begin{proposition}
\label{CharMinPolyofHMatrices}If $Y\in M\left( 2n,\text{ }%
\mathbb{C}
\right) $ is a $\theta _{\mathbb{H}}$-matrix then its minimal and
characteristic polynomials are both real polynomials.
\end{proposition}

\noindent \textbf{Proof:} Let $m_{Y}(x) = x^{k} + c_{k-1}x^{k-1} + \ldots +
c_{0}$

So from 
\begin{equation*}
Y^{k}+c_{k-1}Y^{k-1}+\ldots +c_{1}Y+c_{0}I=0
\end{equation*}%
we get 
\begin{equation*}
(Y^{\ast })^{k}+\bar{c}_{k-1}(Y^{\ast })^{k-1}+\ldots +\bar{c}_{1}Y^{\ast }+%
\bar{c}_{0}I=0
\end{equation*}%
Thus $\bar{m}_{Y}(x)=x^{k}+\bar{c}_{k-1}x^{k-1}+\ldots +\bar{c}_{0}$
annihilates $Y^{\ast }$. Suppose $q(x)=x^{l}+d_{l-1}x^{l-1}+\ldots +d_{0}$
annihilates $Y^{\ast }$, with \underline{$l<k$}. Then the same argument just
used shows that $\bar{q}$, a polynomial of degree $l$, annihilates $Y$. Thus
contradicts the minimality of $m_{Y}(x)$. Hence $k$ is also the degree of
the minimal polynomial of $Y^{\ast }$, and standard properties of minimal
polynomials shows that the minimal polynomial of $Y^{\ast }$ is indeed $\bar{%
m}_{Y}(x)$. But $Y^{\dagger }$ is evidently similar to $Y^{T}$, and thus to $%
Y$. So as $Y$ is a $\theta _{\mathbb{H}}$-matrix, we see that $m_{y}(x)=\bar{%
m}_{Y}(x)$. Hence $m_{Y}(x)$ is a real polynomial.

Next let $p_{Y}(x)=\det (xI-A)$ be the characteristic polynomial of $Y$.
Then the characteristic polynomial of $Y^{\ast }$ is the complex conjugate
of $p_{Y}(\bar{x})$, and hence $p_{Y^{\ast }}(x)=p_{Y}(x)$. But $%
p_{Y^{\dagger }}(x)=p_{Y^{T}}(x)=p_{Y}(x)$. So, as $Y^{\dagger }=Y^{\ast }$,
it is evident that $p_{Y}$ is also a real polynomial. $\diamondsuit $.%
\vspace*{2mm}

Matrices in $sp\left( 4\right) $ are not only $\theta _{\mathbb{H}}$
matrices, but are also anti-Hermitian. This leads to further simplifications
in their minimal polynomials:

\begin{proposition}
\label{shortlist}Let $Y\in sp\left( 4\right) $ and le $m_{Y}(x)$ be its
minimal polynomial. Then $m_{Y}(-x)=m_{Y}(x)$ if the degree of $m_{Y}$ is
even, otherwise $m_{Y}(-x)=-m_{Y}(x)$.
\end{proposition}

\noindent \textbf{Proof:} We have $Y^{\dagger }=-Y$, as $Y\in sp\left(
4\right) $. So the minimal polynomial of $-Y$ is also $m_{Y}$. Hence, if $%
m_{Y}(x)=x^{k}+c_{k-1}x^{k-1}+\ldots +c_{1}x+c_{0}$, it follows that we must
have 
\begin{equation*}
(-Y)^{k}+c_{k-1}(-Y)^{k-1}+\ldots -c_{1}Y+c_{0}I=0
\end{equation*}%
Hence if $k$ is odd, we must have 
\begin{equation*}
Y^{k}-c_{k-1}Y^{k-1}+c_{k-1}Y^{k-2}+\ldots +c_{0}I=0
\end{equation*}%
So $\hat{m}_{Y}(x)=x^{k}-c_{k-1}x^{k-1}+c_{k-1}x^{k-2}+\ldots +c_{0}$ is
also the minimal polynomial of $Y$, and it thus coincides with $m_{Y}(x)$.
This implies that all the even degree terms in $m_{Y}(x)$ vanish.

A similar calculation shows that all the odd degree terms in $m_{Y}(x)$
vanish if $k$ is even. $\diamondsuit$

\begin{remark}
A similar result shows that the characteristic polynomial of $Y\in sp\left(
4\right) $ is a real polynomial with only even degree terms.
\end{remark}

Let us now apply the foregoing results to hone our statements about $m_{Y}(x)
$ for $Y\in sp\left( 4\right) $. Let 
\begin{equation*}
Y=\left( 
\begin{array}{cc}
A & B \\ 
-\bar{B} & \bar{A}%
\end{array}%
\right) 
\end{equation*}

Now $Y\in sp\left( 4\right) $ is equivalent to $(A+Bj)^{\ast }=-(A+Bj)$
(here the $\ast $ is Hermitian conjugation of matrices in $M\left( 2,\text{ }%
\mathbb{H}\right) )$. This is, of course, equivalent to $A^{\ast }=-A$ and $%
B^{T}=B$.

Since the characteristic polynomial of $Y$ is of the form $%
x^{4}+c_{2}x^{2}+c_{0}$, we have 
\begin{equation*}
c_{2}=\frac{1}{2}\{[{\mbox Tr}(Y)]^{2}-{\mbox Tr}(Y^{2})\}
\end{equation*}%
Quite clearly ${\mbox Tr}(Y)=2{\mbox Re}[{\mbox Tr}(A)]$. But as $A$ is
anti-Hermitian its trace is purely imaginary. So ${\mbox Tr}(Y)=0$. Hence 
\begin{equation*}
c_{2}=\frac{-1}{2}{\mbox Tr}(Y^{2})
\end{equation*}%
Now $Y^{2}=\theta _{\mathbb{H}}[(A+Bj)^{2}]$, and 
\begin{equation*}
(A+Bj)^{2}=(A^{2}-B\bar{B})+(AB+B\bar{A})j
\end{equation*}%
Hence 
\begin{equation*}
{\mbox Tr}(Y^{2})=2{\mbox Re}[{\mbox Tr}(A^{2}-B\bar{B})]
\end{equation*}%
But $A^{2}-B\bar{B}=-AA^{\ast }-BB^{\ast }$, which is a negative
semidefinite matrix, and hence a matrix with real trace. So 
\begin{equation*}
c_{2}={\mbox Tr}(AA^{\ast }+BB^{\ast })=\frac{1}{2}\left\Vert Y\right\Vert
_{F}^{2}
\end{equation*}%
So, we have an explicit formula for the characteristic polynomial of $Y$,
viz., 
\begin{equation*}
p_{Y}(x)=x^{4}+(\frac{1}{2}\left\Vert Y\right\Vert _{F}^{2})x^{2}+\det (Y)
\end{equation*}

\begin{remark}
Since the characteristic polynomial of a matrix $Y\in sp\left( 4\right) $ is
a real polynomial on the one hand, and the eigenvalues of $Y$ are purely
imaginary on the other hand, we see that $\det (Y)\geq 0$. Hence $\frac{1}{2}%
\left\Vert Y\right\Vert _{F}^{2}$ is at least as big as the absolute value
of the square root of $\frac{1}{4}\left\Vert Y\right\Vert _{F}^{4}-4\det (Y)$%
, i.e., $\left\Vert Y\right\Vert _{F}^{4}\geq 16\det (Y)$ .
\end{remark}

From this we draw the following conclusions about the eigenstructure of a
non-zero $Y\in sp\left( 4\right) $:

\begin{itemize}
\item $Y$ has $4$ distinct eigenvalues, $ia,$ $-ia,$ $ib,$ $-ib$, iff $%
\left\Vert Y\right\Vert _{F}^{4}>16\det (Y)$ and $\det (Y)\neq 0$.

\item It has $3$ distinct eigenvalues, $ia,$ $-ia,$ $0$ (with $0$ repeated
twice) iff $\det (Y)=0$.

\item It has 2 distinct eigenvalues, $ia$ and $-ia$ (each repeated twice)
iff $\left\Vert Y\right\Vert _{F}^{4}=16\det (Y)$ (notice that in this case $%
Y$ is non-singular, since $Y\neq 0$, precludes $\left\Vert Y\right\Vert
_{F}=0$).
\end{itemize}

Since $Y$ is diagonalizable, the distinct roots of the characteristic
polynomial are the roots, again distinct, of the minimal polynomial. Hence
we find that its minimal polynomials are in each of these cases given as
follows:

\begin{itemize}
\item $x^{4}+(\frac{1}{2}\left\Vert Y\right\Vert _{F}^{2})x^{2}+\det (Y)$.

\item $x^{3}+a^{2}x$. To find $a$, note that the non-zero roots of the
characteristic polynomial are in this case $\frac{i}{\sqrt{2}}\left\Vert
Y\right\Vert _{F},-\frac{i}{\sqrt{2}}\left\Vert Y\right\Vert _{F}$. So $%
a^{2}=\frac{1}{2}\left\Vert Y\right\Vert _{F}^{2}$.

\item $x^{2}+a^{2}$. In this case the roots of the characteristic polynomial
are $\frac{i}{2}\left\Vert Y\right\Vert _{F}$ and $-\frac{i}{2}\left\Vert
Y\right\Vert _{F}$. So the minimal polynomial is $x^{2}+\frac{\left\Vert
Y\right\Vert _{F}^{2}}{4}$.
\end{itemize}

Summarizing we have:

\begin{theorem}
\label{MinPolyForsp4}Let $Y\in sp\left( 4\right) $ or $\widehat{sp}\left(
4\right) $. Its minimal polynomial is one of the following:

\begin{itemize}
\item $x$, which happens iff $Y=0$.

\item $x^{2}+\frac{\left\Vert Y\right\Vert _{F}^{2}}{4}$, which happens iff $%
Y\neq 0$ and $\left\Vert Y\right\Vert _{F}^{4}=16\det (Y)$.

\item $x^{3}+\frac{1}{2}(\left\Vert Y\right\Vert _{F}^{2})x$, which happens
iff $Y\neq 0$, but $\det (Y)=0$.

\item $x^{4}+(\frac{1}{2}\left\Vert Y\right\Vert _{F}^{2})x^{2}+\det (Y)$,
which happens iff $Y\neq 0,$ $\det (Y)\neq 0$.
\end{itemize}
\end{theorem}

\begin{remark}
\label{NoNeedtoConjugateY} Since all quantities intervening in the above
theorem are invariant under real orthogonal similarity, the theorem extends
verbatim to matrices $Y\in \widehat{sp}\left( 4\right) $. Indeed, per Remark
(\ref{ConjugacyHat}), if $Y\in \hat{sp}\left( 4\right) $, then $Z=U^{T}YU$
is in $sp\left( 4\right) $, where $U$ is the explicit real orthogonal matrix
in Remark \ref{ConjugacyHat}. Thus, $i)$ the determinants of $Y$ and $Z$
coincide; $ii)$ $\left\Vert Y\right\Vert _{F}=\left\Vert Z\right\Vert _{F}$;
and $iii)$ the minimal polynomials of $Y$ and $Z$ coincide.
\end{remark}

\begin{remark}
\label{BlockStructureofWideHatsp}Block Structure of $\widehat{sp}\left(
4\right) $: It will be seen that the block structure of a matrix in $%
\widehat{sp}\left( 4\right) $ has some benefits which matrices in $sp\left(
4\right) $ do not. Let $X\in \widehat{sp}\left( 4\right) $. If $X$ is
written as a $2\times 2$ block matrix, with each block $2\times 2$ 
\begin{equation*}
X=\left( 
\begin{array}{cc}
A & B \\ 
C & D%
\end{array}%
\right) 
\end{equation*}%
then $i)$ $A,$ $D$ are both in $sp\left( 2\right) $; $ii)$ $B=-C^{\ast }$
and $iii)$ $B$ is an anti - $\Theta _{\mathbb{H}}$ matrix in $M\left( 2,%
\text{ }\mathbb{%
\mathbb{C}
}\right) $.

To see this, note that $X=\hat{J}_{4}S$ for some $4\times 4$ symmetric
matrix $S=\left( 
\begin{array}{cc}
W & Y \\ 
Y^{T} & Z%
\end{array}%
\right) $ and $X^{\ast }=-X$. Since $\hat{J}_{4}=J_{2}\oplus (-J_{2})$, the
first of these conditions says $A$ and $D$ are in $sp\left( 2,\text{ }%
\mathbb{C}
\right) $, and that  $B=J_{2}Y,$ $C=-J_{2}Y^{T}$ \ Together with the second
condition it follows that $A,$ $D\in sp\left( 2\right) $, and since
$B=-C^{\ast }$
that $Y^{\ast }=J_{2}Y^{T}J_{2}$. This last condition is
equivalent $Y$ being an anti-$\Theta _{\mathbb{H}}$ matrix. Since $B=J_{2}Y$
and $J_{2}$ itself is a $\theta _{\mathbb{H}}$ matrix, it follows that $B$
is an anti-$\Theta _{\mathbb{H}}$ matrix in $M\left( 2,\text{ }\mathbb{%
\mathbb{C}
}\right) $. From this we can conclude the following:

\begin{enumerate}
\item $\left\Vert X\right\Vert _{F}^{2}=2\left( \left\vert x_{11}\right\vert
^{2}+\left\vert x_{12}\right\vert ^{2}+\left\vert x_{33}\right\vert
^{2}+\left\vert x_{34}\right\vert ^{2}\right) +4\left( \left\vert
x_{13}\right\vert ^{2}+\left\vert x_{14}\right\vert ^{2}\right) $

\item The determinant of $X$ requires only the computation of $2\times 2$
determinants. To that end, first observe that an anti - $\theta _{\mathbb{H}}
$ matrix is of the form $\left( 
\begin{array}{cc}
\theta  & \zeta  \\ 
\bar{\zeta} & -\bar{\theta}%
\end{array}%
\right) $, for some $\theta ,$ $\zeta \in \mathbb{%
\mathbb{C}
}$. So it is either invertible or identically zero. Hence, representing $%
X\in \widehat{sp}\left( 4\right) $ as a block matrix, it follows that if $B=0
$, then $\det (X)=\det (A)\det (D)$. If $B$ is invertible, then $\det
(X)=(-1)^{4}\det (B)\det (-B^{\ast }-DB^{-1}A)=\det (B)\det (B^{\ast
}+DB^{-1}A)$, which follows from the special case of the determinantal
formulae of Schur mentioned in Section $2.7$.
\end{enumerate}

The last item above shows that for a determinant calculation at least $%
\widehat{sp}\left( 4\right) $ is more amenable than $sp\left( 4\right) $.
Indeed, if $\left( 
\begin{array}{cc}
A & B \\ 
-\bar{B} & \bar{A}%
\end{array}%
\right) \in sp\left( 4\right) $, then one will need a $4\times 4$
determinant calculation, when both $A$ and $B$ fail to be invertible, since
it is now possible for $A$ and $B$ to be singular without being identically
zero.
\end{remark}

\begin{remark}
There is an alternative characterization of when $Y\in sp\left( 4\right) $
possesses a quadratic minimal polynomial. This characterization is mostly
applicable for $Y\in sp\left( 2n\right) $ also. Consider $Y=\theta (A+Bj)\in
sp\left( 2n\right) $. Squaring $Y$, we find 
\begin{equation*}
Y^{2}=\theta _{\mathbb{H}}[(A+Bj)^{2}]=\theta _{\mathbb{H}}[(-A^{2}-B\bar{B}%
)+(AB+B\bar{A})j]
\end{equation*}%
But $A^{\ast }=-A$ and $\bar{B}=B^{\ast }$. Similarly $\bar{A}=-A^{T}$,
while $B=B^{T}$. So we find 
\begin{equation*}
Y^{2}=\theta _{\mathbb{H}}[(-AA^{\ast }-BB^{\ast }))+(AB-(AB)^{T})j]
\end{equation*}%
So $Y^{2}=-c^{2}Y$ for some $c\in 
\mathbb{R}
$, iff the positive semidefinite matrix $AA^{\ast }+BB^{\ast }$ is a scalar
matrix, and the matrix $AB$ is symmetric.

Now these 2 conditions are also equivalent to $A+Bj$ being, upto a positive
constant, an unitary element of $M\left( 2,\text{ }\mathbb{H}\right) $,
i.e., to $(A+Bj)(A+Bj)^{\ast }=c^{2}I_{2}$, for some $c\in 
\mathbb{R}
$. Indeed%
\begin{equation*}
(A+Bj(A+Bj)^{\ast }=(AA^{\ast }+BB^{\ast }+(BA^{T}-AB^{T})j
\end{equation*}%
Once again, using $B^{T}=B$, we conclude that 
\begin{equation*}
(A+Bj)(A+Bj)^{\ast }=(AA^{\ast }+BB^{\ast })+((AB)^{T}-(AB))j=c^{2}I
\end{equation*}

When $n=2$, these lead to easily verfied conditions on the entries of $A$
and $B$. Specifically, if $A=\left( 
\begin{array}{cc}
ia & z_{1} \\ 
-\bar{z_{1}} & ib%
\end{array}%
\right) $ and $B=\left( 
\begin{array}{cc}
z_{2} & z_{3} \\ 
z_{3} & z_{4}%
\end{array}%
\right) $, then $Y=\theta _{\mathbb{H}}(A+Bj)$ has a quadratic minimal
polynomial iff 
\begin{eqnarray*}
a^{2}+\left\vert z_{2}\right\vert ^{2} &=&b^{2}+\left\vert z_{4}\right\vert
^{2} \\
\bar{z}_{3}z_{2}+z_{3}\bar{z}_{4} &=&ia\bar{z}_{1}+ibz_{1} \\
z_{1}(z_{4}+z_{2}) &=&i(b-a)z_{3}
\end{eqnarray*}

One can write down conditions on $A$ and $B$ for an arbitrary $Y=\theta _{%
\mathbb{H}}(A+Bj)$ in $sp\left( 2n\right) $ to have $x^{3}+c^{2}x$ as its
minimal polynomial, by directly computing $(A+Bj)^{3}$. However, these
conditions don't lead to any succinctly stated conditions even when $n=2$.
\end{remark}

\section{$\mathfrak{su}\left( 4\right) $ and $\mathfrak{so}\left( 6,\text{ }%
\mathbb{R}
\right) $}

As is well known the spin group of $SO\left( 6,\text{ }%
\mathbb{R}
\right) $ is $SU\left( 4\right) $, and there is correspondingly an
isomorphism of $\mathfrak{so}\left( 6,\text{ }%
\mathbb{R}
\right) $ and $\mathfrak{su}\left( 4\right) $. In this section we will
produce a basis of $1$-vectors of ${\mbox C}l\left( 0,\text{ }6\right) $
which is natural from the point of view of the constructions of Sec $2.3$
and which will enable the computation of exponentials of matrices in $%
\mathfrak{so}\left( 6,\text{ }%
\mathbb{R}
\right) $ via a computation of exponentials of matrices in $\mathfrak{su}%
\left( 4\right) $. Moreover in this construction, the matrix $\widetilde{J}%
_{8}$ naturally intervenes.

We begin with ${\mbox C}l\left( 0,\text{ }0\right) $ and repeatedly apply 
\textbf{IC1} of Sec $2.3$, to first produce a basis of $1$-vectors for ${%
\mbox C}l\left( 3,\text{ }3\right) =M\left( 8,\text{ }%
\mathbb{R}
\right) $.

Since the set of $1$-vectors for ${\mbox C}l\left( 0,\text{ }0\right) $ is
the empty set, $\{\sigma _{x},$ $\sigma _{y}\}$ is what \textbf{IC1} gives
for a basis of $1$-vectors for ${\mbox C}l(1,1)$.

Hence a basis of $1$-vectors for ${\mbox C}l(2,2)$ is then 
\begin{equation*}
\left( 
\begin{array}{cc}
\sigma _{x} & 0 \\ 
0 & -\sigma _{x}%
\end{array}%
\right) ;\text{ }\left( 
\begin{array}{cc}
0 & I_{2} \\ 
I_{2} & 0%
\end{array}%
\right) ;\text{ }\left( 
\begin{array}{cc}
i\sigma _{y} & 0 \\ 
0 & -i\sigma _{y}%
\end{array}%
\right) ;\text{ }\left( 
\begin{array}{cc}
0 & I_{2} \\ 
-I_{2} & 0%
\end{array}%
\right) 
\end{equation*}%
This produces the following basis of $1$-vectors for ${\mbox C}l\left( 3,%
\text{ }3\right) $%
\begin{equation*}
\left\{ \sigma _{z}\otimes \sigma _{z}\otimes \sigma _{x},\text{ }\sigma
_{z}\otimes \sigma _{x}\otimes I_{2}\sigma _{z}\otimes \sigma _{z}\otimes
i\sigma _{y},\text{ }\sigma _{z}\otimes i\sigma _{y}\otimes I_{2},\text{ }%
\sigma _{x}\otimes I_{4},\text{ }i\sigma _{y}\otimes I_{4}\right\} 
\end{equation*}

Next, we use \textbf{IC3} of Sec $2.3$, relating ${\mbox C}l\left( p,\text{ }%
q\right) $ and ${\mbox C}l(p+1,q-1)$, to produce, via this basis, a basis of 
$1$-vectors for ${\mbox C}l\left( 4,\text{ }2\right) $:

\begin{center}
\begin{tabular}{lllllll}
$\tilde{e}_{1}$ & $=$ & $\sigma _{z}\otimes \sigma _{z}\otimes \sigma _{x}$
&  & $\tilde{e}_{4}$ & $=$ & $(i\sigma _{y}\otimes I_{4})(\sigma _{z}\otimes
\sigma _{z}\otimes \sigma _{x})$ \\ 
$\tilde{e}_{2}$ & $=$ & $(\sigma _{z}\otimes \sigma _{z}\otimes i\sigma
_{y})(\sigma _{z}\otimes \sigma _{z}\otimes \sigma _{x})$ &  & $\tilde{e}_{5}
$ & $=$ & $(\sigma _{z}\otimes \sigma _{x}\otimes I_{2})(\sigma _{z}\otimes
\sigma _{z}\otimes \sigma _{x})$ \\ 
$\tilde{e}_{3}$ & $=$ & $(\sigma _{z}\otimes i\sigma _{y}\otimes
I_{2})(\sigma _{z}\otimes \sigma _{z}\otimes \sigma _{x})$ &  & $\tilde{e}%
_{6}$ & $=$ & $(\sigma _{x}\otimes I_{4})(\sigma _{z}\otimes \sigma
_{z}\otimes \sigma _{x})$%
\end{tabular}
\end{center}

Doing the requisite Kronecker multiplications this basis of $1$-vectors for $%
{\mbox C}l\left( 4,\text{ }2\right) $ assumes the following form:

\begin{center}
\begin{tabular}{lllllll}
$\tilde{e}_{1}$ & $=$ & $\sigma _{z}\otimes \sigma _{z}\otimes \sigma _{x}$
&  & $\tilde{e}_{4}$ & $=$ & $-\sigma _{x}\otimes \sigma _{z}\otimes \sigma
_{x}$ \\ 
$\tilde{e}_{2}$ & $=$ & $I_{2}\otimes I_{2}\otimes \sigma _{z}$ &  & $\tilde{%
e}_{5}$ & $=$ & $-I_{2}\otimes i\sigma _{y}\otimes \sigma _{x}$ \\ 
$\tilde{e}_{3}$ & $=$ & $-I_{2}\otimes \sigma _{x}\otimes \sigma _{x}$ &  & $%
\tilde{e}_{6}$ & $=$ & $-i\sigma _{y}\otimes \sigma _{z}\otimes \sigma _{x}$%
\end{tabular}
\end{center}

Finally, using \textbf{IC2} of Sec $2.3$, relating ${\mbox C}l\left( p,\text{
}q\right) $ to ${\mbox C}l(p-4,q+4)$, produces a basis of $1$-vectors for $%
{\mbox C}l\left( 0,\text{ }6\right) $. To that end, we first need to find $%
\tilde{e}_{1}\tilde{e}_{2}\tilde{e}_{3}\tilde{e}_{4}$. This is given by 
\begin{equation*}
\tilde{e}_{1}\tilde{e}_{2}\tilde{e}_{3}\tilde{e}_{4}=i\sigma _{y}\otimes
\sigma _{x}\otimes i\sigma _{y}
\end{equation*}

This results in a  basis of $1$-vectors 
for ${\mbox C}l\left( 0,\text{ }6\right) $ as shown in Table 4.

\begin{table}[tbp] \centering%
$%
\begin{tabular}{lllll}
$Y_{1}$ & $=$ & $(\sigma _{z}\otimes \sigma _{z}\otimes \sigma _{x})(i\sigma
_{y}\otimes \sigma _{x}\otimes i\sigma _{y})$ & $=$ & $\sigma _{x}\otimes
(i\sigma _{y})\otimes (-\sigma _{z})$ \\ 
$Y_{2}$ & $=$ & $(I_{2}\otimes I_{2}\otimes \sigma _{z})(i\sigma _{y}\otimes
\sigma _{x}\otimes i\sigma _{y})$ & $=$ & $i\sigma _{y}\otimes \sigma
_{x}\otimes \sigma _{x}$ \\ 
$Y_{3}$ & $=$ & $(-I_{2}\otimes \sigma _{x}\otimes \sigma _{x})(i\sigma
_{y}\otimes \sigma _{x}\otimes i\sigma _{y})$ & $=$ & $i\sigma _{y}\otimes
I_{2}\otimes \sigma _{z}$ \\ 
$Y_{4}$ & $=$ & $(-\sigma _{x}\otimes \sigma _{z}\otimes \sigma
_{x})(i\sigma _{y}\otimes \sigma _{x}\otimes i\sigma _{y})$ & $=$ & $-\sigma
_{z}\otimes (i\sigma _{y})\otimes \sigma _{z}$ \\ 
$Y_{5}$ & $=$ & $-I_{2}\otimes i\sigma _{y}\otimes \sigma _{x}$ & $=$ & $%
-I_{2}\otimes i\sigma _{y}\otimes \sigma _{x}$ \\ 
$Y_{6}$ & $=$ & $-i\sigma _{y}\otimes \sigma _{z}\otimes \sigma _{x}$ & $=$
& $-i\sigma _{y}\otimes \sigma _{z}\otimes \sigma _{x}$%
\end{tabular}%
$\caption{Basis of $1$-vectors for ${\mbox C}l(0,6)$}\label{TableKey_3}%
\end{table}%

\begin{remark}
\label{Cliffon06}Each of the $Y_{i}$ are tensor products of $3$ matrices, of
which two are real symmetric and one is real antisymmetric. Hence, $%
Y_{i}^{T}=-Y_{i},$ for all $i$. Since matrix transposition is an
anti-involution, we find, as expected, from this that (with respect to this
basis of $1$-vectors), Clifford conjugation on ${\mbox C}l\left( 0,\text{ }%
6\right) $ coincides with matrix transposition.
\end{remark}

Next a matrix form for reversion on ${\mbox C}l\left( 0,\text{ }6\right) $
(with respect to the basis, $\left\{ Y_{i}\mid i=1,\ldots ,6\right\} $, of $1
$-vectors) will be found. We are guided in this by 3 facts: i) the $Y_{i}$
are all tensor products of $3$ matrices, and the matrix $i\sigma _{y}$ is
one of the 3 factors in each $Y_{i}$; ii) the matrices $J_{8}$ and $\tilde{J}%
_{8}$ are also triple tensor products with $i\sigma _{y}$ again one of the
factors. Specifically, $J_{8}=i\sigma _{y}\otimes I_{4}=i\sigma _{y}\otimes
I_{2}\otimes I_{2}$ and $\tilde{J}_{8}=I_{4}\otimes (i\sigma
_{y})=I_{2}\otimes I_{2}\otimes (i\sigma _{y})$; and iii) Neither $J_{8}$
nor $\tilde{J}_{8}$ are any of the $Y_{i},i=1,\ldots ,6$. In view of the
multiplication table for the Pauli matrices, it is natural to seek reversion
in the form $M^{-1}X^{T}M$, with $M$ either $J_{8}$ or $\widetilde{J}_{8}$.
A few calculations reveal that $J_{8}^{-1}Y_{i}^{T}J_{8}\neq Y_{i},\forall i$%
. Hence, reversion cannot be given by $J_{8}^{-1}X^{T}J_{8}$. However, we
have the following proposition:

\begin{proposition}
\label{Reversionon06}

\begin{description}
\item[i)] The reversion anti-involution on ${\mbox C}l\left( 0,\text{ }%
6\right) $, with respect to the basis%
\begin{equation*}
\begin{tabular}{lllllll}
$Y_{1}$ & $=$ & $\sigma _{x}\otimes (i\sigma _{y})\otimes (-\sigma _{z})$ & 
& $Y_{4}$ & $=$ & $-\sigma _{z}\otimes (i\sigma _{y})\otimes \sigma _{z}$ \\ 
$Y_{2}$ & $=$ & $i\sigma _{y}\otimes \sigma _{x}\otimes \sigma _{x}$ &  & $%
Y_{5}$ & $=$ & $-I_{2}\otimes i\sigma _{y}\otimes \sigma _{x}$ \\ 
$Y_{3}$ & $=$ & $i\sigma _{y}\otimes I_{2}\otimes \sigma _{z}$ &  & $Y_{6}$
& $=$ & $-i\sigma _{y}\otimes \sigma _{z}\otimes \sigma _{x}$%
\end{tabular}%
\end{equation*}

of $1$-vectors is given by $\Phi ^{rev}(X)=\widetilde{J}_{8}^{T}X^{T}%
\widetilde{J}_{8},$ for all $X\in {\mbox C}l\left( 0,\text{ }6\right) $.

\item[ii)] The grade involution on ${\mbox C}l\left( 0,\text{ }6\right) $,
with respect to the basis $\left\{ Y_{i}\mid i=1,\ldots ,6\right\} $ of $1$%
-vectors is given by $\Phi ^{gr}(X)=\widetilde{J}_{8}^{T}X\widetilde{J}_{8}$%
. Thus, the algebra of even vectors in ${\mbox C}l\left( 0,\text{ }6\right) $
is the image of $M\left( 4,\text{ }%
\mathbb{C}
\right) $, under $\theta _{%
\mathbb{C}
}$, in $M\left( 8,\text{ }%
\mathbb{R}
\right) $.
\end{description}
\end{proposition}

\noindent \textbf{Proof:} First note that 
\begin{equation*}
\widetilde{J}_{8}^{-1}=\widetilde{J}_{8}^{T}=I_{2}\otimes I_{2}\otimes
(-i\sigma _{Y})
\end{equation*}%
Next, it suffices to to check that the map $X\rightarrow \widetilde{J}%
_{8}^{T}X^{T}\widetilde{J}_{8}$, which is evidently an anti-involution, is
the identity map on $1$-vectors. For this, in turn, it suffices to verify
that $\widetilde{J}_{8}^{T}Y_{i}^{T}\widetilde{J}_{8}=Y_{i},$ for all $%
i=1,\ldots ,6$. This computation is facilitated by the representations of
the $Y_{i},\widetilde{J}_{8},\widetilde{J}_{8}^{T}$ all as threefold
Kronecker products. We will content ourselves with demonstrating this for $%
Y_{1}$: 
\begin{equation*}
\widetilde{J}_{8}^{T}Y_{1}^{T}\widetilde{J}_{8}=[I_{2}\otimes I_{2}\otimes
(-i\sigma _{y})][\sigma _{x}\otimes (i\sigma _{y})\otimes (-\sigma
_{z})]^{T}[(I_{2}\otimes I_{2}\otimes (i\sigma _{y})]
\end{equation*}%
Using the fact that $i\sigma _{y}$ is antisymmetric, while $\sigma
_{x},\sigma _{z}$ are symmetric, we find that $\widetilde{J}_{8}^{T}Y_{1}^{T}%
\widetilde{J}_{8}$, is therefore 
\begin{equation*}
\lbrack I_{2}\otimes I_{2}\otimes (-i\sigma _{y})][\sigma _{x}\otimes
(-i\sigma _{y})\otimes (-\sigma _{z})][(I_{2}\otimes I_{2}\otimes (i\sigma
_{y})]=\sigma _{x}\otimes (i\sigma _{y})\otimes (-\sigma _{z})=Y_{1}
\end{equation*}

A similar computation reveals the result to hold for the remaining $Y_{i}$'s.

The second part of the proposition now is just a consequence of of the last
sentence of Remark \ref{Cliffon06}. Hence, being an even vector is
equivalent to $X=\widetilde{J}_{8}^{T}X\widetilde{J}_{8}$, i.e., to $X^{T}=%
\tilde{J}_{8}^{T}X^{T}\tilde{J}_{8}$, which by $v)$ of Remark \ref%
{Cproperties} says precisely that $X=\Theta _{%
\mathbb{C}
}(Y)$ for some $Y\in M\left( 4,\text{ }%
\mathbb{C}
\right) $. $\diamondsuit $\vspace*{3mm}

This yields the following:

\begin{corollary}
Consider the basis of $1$-vectors for ${\mbox C}l(1,7)$ given by $\left\{
K_{16},\text{ }\left( 
\begin{array}{cc}
Y_{i} & 0 \\ 
0 & -Y_{i}%
\end{array}%
\right) ,\text{ }J_{16}\right\} $, where $Y_{i},$ $i=1,\ldots ,6$ is as in
Proposition \ref{Reversionon06}. Then for $X\in {\mbox C}l(1,7)=M(16,%
\mathbb{R}
)$, the following hold:

\begin{itemize}
\item $X^{cc}=Q^{-1}X^{T}Q$, with $Q=\left( 
\begin{array}{cc}
0_{8} & \widetilde{J}_{8} \\ 
\widetilde{J}_{8} & 0_{8}%
\end{array}%
\right) $.

\item $X^{rev}=K_{16}^{-1}X^{T}K_{16}$.
\end{itemize}
\end{corollary}

Returning to ${\mbox C}l\left( 0,\text{ }6\right) $, it now follows that ${%
\mbox Spin}(6)$ is the collection of $Z\in {\mbox C}l\left( 0,\text{ }%
6\right) =M\left( 8,\text{ }%
\mathbb{R}
\right) $ satsifying

\begin{description}
\item[i)] $ZZ^{T}=I_{n}$.

\item[ii)] $Z$ is even, i.e., $Z=\Theta _{%
\mathbb{C}
}(W)$, for some $W\in M\left( 4,\text{ }%
\mathbb{C}
\right) $.

\item[iii)] $ZYZ^{T}$ is a $1$-vector for all $1$-vectors $Y\in {\mbox C}%
l\left( 0,\text{ }6\right) $.
\end{description}

The first two conditions say that $Z=\Theta _{%
\mathbb{C}
}(W)$ for some $W\in U\left( 4\right) $. However, as is well known, unlike
the case of ${\mbox Spin}\left( 5\right) $, the last condition is no longer
superfluous. Dimension considerations say that the third condition forces the
corresponding $W$ to be a connected 15 dimensional subgroup of $U\left(
4\right) $. The obvious candidate is $SU\left( 4\right) $. Within the
context of the derivation above, this can be verified in one of several
explicit ways. For instance,

\begin{description}
\item[I)] Suppose we have a set of generators $M_{k}$ for $SU\left( 4\right) 
$, i.e., every element of $SU\left( 4\right) $ can be factorized into a
product of the $M_{k}$'s. Then it suffices to check that $\theta _{%
\mathbb{C}
}(M_{k})Y_{i}[\theta _{%
\mathbb{C}
}(M_{k})]^{-1}$ is a real linear combination of the $Y_{i}$'s for each $%
Y_{i}$ and each $M_{k}$. Here, as before $\{Y_{i}\},i=1,\ldots ,6$ is the
basis of $1$-vectors of ${\mbox C}l\left( 0,\text{ }6\right) $ in
Proposition \ref{Reversionon06}. Given the Kronecker product representations
of the $Y_{i}$, a convenient choice for the $M_{k}$ is the following
collection of matrices 
\begin{eqnarray*}
&&\{I_{2}\otimes \left( \exp \left[ i\left( \alpha \sigma _{x}+\beta \sigma
_{y}+\gamma \sigma _{z}\right) \right] \right) ,\text{ }\left( \exp \left[
i\left( \mu \sigma _{x}+\nu \sigma _{y}+\eta \sigma _{z}\right) \right]
\right) \otimes I_{2}, \\
&&\exp \left( ia\sigma _{x}\otimes \sigma _{x}\right) ,\text{ }\exp \left(
ib\sigma _{y}\otimes \sigma _{y}\right) ,\text{ }\exp \left( ic\sigma
_{z}\otimes \sigma _{z}\right) \}
\end{eqnarray*}%
Here, $\alpha ,\beta ,\gamma ,\mu ,\nu ,\eta ,a,b,c\in 
\mathbb{R}
$. This is one of the so-called \textit{KAK} decompositions of $SU\left(
4\right) $ and is very useful in quantum information theory, for instance.

\item[II)] For each element $X$ of a basis for $\mathfrak{su}\left( 4\right) 
$, it suffices to check $\theta _{%
\mathbb{C}
}(X)Y_{i}-Y_{i}\theta _{%
\mathbb{C}
}(X)$ is a real linear combination of the $Y_{i}$'s.
\end{description}

Verification of item $II)$ is carried out in Theorem \ref{su4so6Isomorphism}
below, since it will be needed at other points as well. It is also
interesting to note that the archtypal element in the Lie algebra $u\left(
4\right) $, \underline{but not in $\mathfrak{su}\left( 4\right) $}, viz., $%
iI_{4}$, violates the linearization of the third condition for ${\mbox Spin}%
(6)$ in a rather strong way. In other words, denoting by $V$, the matrix $%
I_{4}\otimes (i\sigma _{y})=\Theta _{%
\mathbb{C}
}(iI_{4})$, one finds that $VY_{i}-Y_{i}V$ is not a $1$-vector for 
\underline{any} $Y_{i}$. We will just demonstrate this for $Y_{1}$.
Computing $VY_{1}-Y_{1}V$, we find that it equals 
\begin{equation*}
(I_{2}\otimes I_{2}\otimes (i\sigma _{y})(\sigma _{x}\otimes (i\sigma
_{y})\otimes (-\sigma _{z})-(\sigma _{x}\otimes (i\sigma _{y})\otimes
(-\sigma _{z})(I_{2}\otimes I_{2}\otimes (i\sigma _{y})=2\sigma _{x}\otimes
(i\sigma _{y})\otimes \sigma _{x}
\end{equation*}%
If we denote the end product of this computation by $\Lambda _{1}$, then $%
\Lambda _{1}$ is, in fact, orthogonal to every $1$-vector, with respect to
the trace inner product on $M\left( 8,\text{ }%
\mathbb{R}
\right) ={\mbox C}l\left( 0,\text{ }6\right) $. This is because a quick
calculation of the matrices $\Lambda _{1}^{T}Y_{i}$ reveals that each of
them is a threefold Kronecker product, in which at least one factor is a
multiple of one of the Pauli matrices $\sigma _{i},i=x,y,z$. Since the Pauli
matrices are traceless, it follows that each $\Lambda _{1}^{T}Y_{i}$ is
traceless. Similar calculations show that $VY_{i}-Y_{i}V$ is not a $1$%
-vector for $i\geq 2$ also.

On the other hand, the calculations below confirm that if $V=\theta _{%
\mathbb{C}
}(W),W\in \mathfrak{su}\left( 4\right) $, then $VY_{i}-Y_{i}V$ is a $1$%
-vector, $\forall i=1,\ldots ,6$.\vspace*{3mm}

\noindent \textbf{Computation of the Lie Algebra Isomorphism Between $%
\mathfrak{su}\left( 4\right) $ and $\mathfrak{so}$}$\left( 6,\text{ }%
\mathbb{R}
\right) $\textbf{:}

To achieve the said computation we first need to identify the elements of $%
M\left( 8,\text{ }%
\mathbb{R}
\right) $ which arise as $\Theta _{%
\mathbb{C}
}(X)$, as $X$ runs over a basis of $\mathfrak{su}\left( 4\right) $. The
basis of $\mathfrak{su}\left( 4\right) $ we will work with is the basis
consisting of Kronecker products of the Pauli matrices (including $\sigma
_{0}=I_{2})$. We then obtain Table \ref{ThetaCsu4}. 

\begin{table}[tbp] \centering%
\begin{tabular}{llllllll}
$X\in \mathfrak{su}\left( 4\right) $ & $\Theta _{%
\mathbb{C}
}(X)$ &  & $X\in \mathfrak{su}\left( 4\right) $ & $\Theta _{%
\mathbb{C}
}(X)$ &  & $X\in \mathfrak{su}\left( 4\right) $ & $\Theta _{%
\mathbb{C}
}(X)$ \\ 
$i\sigma _{x}\otimes I_{2}$ & $\sigma _{x}\otimes I_{2}\otimes (i\sigma _{y})
$ &  & $I_{2}\otimes (i\sigma _{z})$ & $I_{2}\otimes \sigma _{x}\otimes
(i\sigma _{y})$ &  & $i\sigma _{x}\otimes \sigma _{y}$ & $\sigma _{x}\otimes
(i\sigma _{y})\otimes I_{2}$ \\ 
$i\sigma _{y}\otimes I_{2}$ & $i\sigma _{y}\otimes I_{2}\otimes I_{2}$ &  & 
$i\sigma _{z}\otimes \sigma _{z}$ & $\sigma _{z}\otimes \sigma _{z}\otimes
(i\sigma _{y})$ &  & $i\sigma _{x}\otimes \sigma _{z}$ & $\sigma _{x}\otimes
\sigma _{z}\otimes (i\sigma _{y})$ \\ 
$i\sigma _{z}\otimes I_{2}$ & $\sigma _{z}\otimes I_{2}\otimes (i\sigma _{y})
$ &  & $i\sigma _{z}\otimes \sigma _{x}$ & $\sigma _{z}\otimes \sigma
_{z}\otimes (i\sigma _{y})$ &  & $i\sigma _{y}\otimes \sigma _{x}$ & $%
i\sigma _{y}\otimes \sigma _{x}\otimes I_{2}$ \\ 
$I_{2}\otimes (i\sigma _{x})$ & $I_{2}\otimes \sigma _{x}\otimes (i\sigma
_{y})$ &  & $i\sigma _{z}\otimes \sigma _{y}$ & $\sigma _{z}\otimes (i\sigma
_{y})\otimes I_{2}$ &  & $i\sigma _{y}\otimes \sigma _{y}$ & $i\sigma
_{y}\otimes (i\sigma _{y})\otimes (i\sigma _{y})$ \\ 
$I_{2}\otimes (i\sigma _{y})$ & $I_{2}\otimes (i\sigma _{y})\otimes I_{2}$ & 
& $i\sigma _{x}\otimes \sigma _{x}$ & $\sigma _{x}\otimes \sigma _{x}\otimes
(i\sigma _{y})$ &  & $i\sigma _{y}\otimes \sigma _{z}$ & $i\sigma
_{y}\otimes \sigma _{z}\otimes I_{2}$%
\end{tabular}%
\caption{ $\Theta _{\mathbb{C}}$ embedding of $\mathfrak{su}\left( 4\right)$
}\label{ThetaCsu4}%
\end{table}%

We can now state:

\begin{theorem}
\label{su4so6Isomorphism}The Lie algebra isomorphism $\Psi _{6}:\mathfrak{su}%
\left( 4\right) \rightarrow \mathfrak{so}\left( 6,\text{ }%
\mathbb{R}
\right) $ is prescribed by its effect on the basis $\{i\sigma _{j}\otimes
I_{2},$ $I_{2}\otimes (i\sigma _{k}),$ $i\sigma _{p}\otimes \sigma _{q}\},$ $%
j,$ $k,$ $p,$ $q\in \{x,$ $y,$ $z\}$ of $\mathfrak{su}\left( 4\right) $ via
Table \ref{ThetaCsu5_1}$. $ 
\begin{table}[tbp] \centering%
\begin{tabular}{llllll}
Basis of $\mathfrak{su}\left( 4\right) $ & \hspace{-0.1in}Basis of $%
\mathfrak{so}\left( 6,\text{ }%
\mathbb{R}
\right) $ & Basis of $\mathfrak{su}\left( 4\right) $ & \hspace{-0.1in}Basis
of $\mathfrak{so}\left( 6,\text{ }%
\mathbb{R}
\right) $ & Basis of $\mathfrak{su}\left( 4\right) $ & \hspace{-0.1in}Basis
of $\mathfrak{so}\left( 6,\text{ }%
\mathbb{R}
\right) $ \\ 
$i\sigma _{x}\otimes I_{2}$ & \hspace{-0.1in}$%
2(e_{1}e_{5}^{T}-e_{5}e_{1}^{T})$ & $I_{2}\otimes (i\sigma _{z})$ & \hspace{%
-0.1in}$2(e_{6}e_{3}^{T}-e_{3}e_{6}^{T})$ & $i\sigma _{x}\otimes \sigma _{y}$
& \hspace{-0.1in}$2(e_{4}e_{3}^{T}-e_{3}e_{4}^{T})$ \\ 
$i\sigma _{y}\otimes I_{2}$ & \hspace{-0.1in}$%
2(e_{4}e_{1}^{T}-e_{1}e_{4}^{T})$ & $i\sigma _{z}\otimes \sigma _{z}$ & 
\hspace{-0.1in}$2(e_{2}e_{1}^{T}-e_{1}e_{2}^{T})$ & $i\sigma _{x}\otimes
\sigma _{z}$ & \hspace{-0.1in}$2(e_{4}e_{2}^{T}-e_{2}e_{4}^{T})$ \\ 
$i\sigma _{z}\otimes I_{2}$ & \hspace{-0.1in}$%
2(e_{4}e_{5}^{T}-e_{5}e_{4}^{T})$ & $i\sigma _{z}\otimes \sigma _{x}$ & 
\hspace{-0.1in}$2(e_{6}e_{1}^{T}-e_{1}e_{6}^{T})$ & $i\sigma _{y}\otimes
\sigma _{x}$ & \hspace{-0.1in}$2(e_{5}e_{6}^{T}-e_{6}e_{5}^{T})$ \\ 
$I_{2}\otimes (i\sigma _{x})$ & \hspace{-0.1in}$%
2(e_{3}e_{2}^{T}-e_{2}e_{3}^{T})$ & $i\sigma _{z}\otimes \sigma _{y}$ & 
\hspace{-0.1in}$2(e_{3}e_{1}^{T}-e_{1}e_{3}^{T})$ & $i\sigma _{y}\otimes
\sigma _{y}$ & \hspace{-0.1in}$2(e_{3}e_{5}^{T}-e_{5}e_{3}^{T})$ \\ 
$I_{2}\otimes (i\sigma _{y})$ & \hspace{-0.1in}$%
2(e_{2}e_{6}^{T}-e_{6}e_{2}^{T})$ & $i\sigma _{x}\otimes \sigma _{x}$ & 
\hspace{-0.1in}$2(e_{4}e_{6}^{T}-e_{6}e_{4}^{T})$ & $i\sigma _{y}\otimes
\sigma _{z}$ & \hspace{-0.1in}$2(e_{2}e_{5}^{T}-e_{5}e_{2}^{T})$%
\end{tabular}%
\caption{Lie algebra isomorphism between $\mathfrak{su}\left( 4\right)$ and
$\mathfrak{so}\left( 6,\text{ }\mathbb{R}\right) $ }\label{ThetaCsu5_1}%
\end{table}%
\end{theorem}

\noindent \textbf{Proof:} Let us label each of the matrices displayed 
in the
II column of Table \ref{ThetaCsu4} as $A_{k}, k=1, \ldots , 15$.
(for instance, $A_{2} =i\sigma_{y}\otimes I_{2}\otimes I_{2}$).
For each such $A_{k}$, we compute $A_{k}Y_{i}-Y_{i}A_{k}$,
where $\{Y_{1},\ldots ,Y_{6}\}$ is the basis of $1$-vectors of ${\mbox C}%
l\left( 0,\text{ }6\right) $ and express the result as a linear combination
of the $Y_{l},$ $l=1,\ldots ,6$. The resulting matrix is the image of $\Psi
_{6}(X)$, where $X$ is an element of the basis of $\mathfrak{su}\left(
4\right) $ listed in the I column of Table \ref{ThetaCsu4}. This is a long
calculation. We will just record the details for $A_{2}$ for illustration.
We compute

\noindent 
\begin{tabular}{lll}
$A_{2}Y_{1}-Y_{1}A_{2}$ & $=$ & $(i\sigma _{y}\otimes I_{2}\otimes
I_{2})(\sigma _{x}\otimes (i\sigma _{y}\otimes (-\sigma _{z})$ \\ 
&  & $-(\sigma _{x}\otimes (i\sigma _{y}\otimes (-\sigma _{z})(i\sigma
_{y}\otimes I_{2}\otimes I_{2})$ \\ 
& $=$ & $-2\sigma _{z}\otimes (i\sigma _{y})\otimes \sigma _{z}=2Y_{4}$%
\end{tabular}%
\bigskip

\noindent 
\begin{tabular}{lll}
$A_{2}Y_{2}-Y_{2}A_{2}$ & $=$ & $(i\sigma _{y}\otimes I_{2}\otimes
I_{2})(i\sigma _{y}\otimes \sigma _{x}\otimes \sigma _{x})$ \\ 
&  & $-(i\sigma _{y}\otimes \sigma _{x}\otimes \sigma _{x})(i\sigma
_{y}\otimes I_{2}\otimes I_{2})=0$%
\end{tabular}%
\bigskip

\noindent 
\begin{tabular}{lll}
$A_{2}Y_{5}-Y_{5}A_{2}$ & $=$ & $i\sigma _{y}\otimes I_{2}\otimes
I_{2})(-I_{2}\otimes (i\sigma _{y})\otimes \sigma _{x})$ \\ 
&  & $-(\sigma _{z}\otimes (i\sigma _{y}\otimes (-\sigma _{z})(i\sigma
_{y}\otimes I_{2}\otimes I_{2})$ \\ 
& $=$ & $2\sigma _{x}\otimes (i\sigma _{y})\otimes \sigma _{z}=-2Y_{1}$%
\end{tabular}%
\bigskip

\noindent 
\begin{tabular}{lll}
$A_{2}Y_{5}-Y_{5}A_{2}$ & $=$ & $i\sigma _{y}\otimes I_{2}\otimes
I_{2})(-I_{2}\otimes (i\sigma _{y})\otimes \sigma _{x})$ \\ 
&  & $-(-I_{2}\otimes (i\sigma _{y})\otimes \sigma _{x})(i\sigma _{y}\otimes
I_{2}\otimes I_{2})=0$%
\end{tabular}%
\bigskip

\noindent 
\begin{tabular}{lll}
$A_{2}Y_{6}-Y_{6}A_{2}$ & $=$ & $i\sigma _{y}\otimes I_{2}\otimes
I_{2})(-i\sigma _{y}\otimes \sigma _{z}\otimes \sigma _{x})$ \\ 
&  & $-(-i\sigma _{y}\otimes \sigma _{z}\otimes \sigma _{x})(i\sigma
_{y}\otimes I_{2}\otimes I_{2})=0$%
\end{tabular}%
\smallskip

\noindent Hence $\Psi_{6} (i\sigma _{y}\otimes
I_{2})=2(e_{4}e_{1}^{T}-e_{1}e_{4}^{T})\smallskip $

\noindent \textit{Computing Exponentials in }$\mathfrak{so}\left( 6,\text{ }%
\mathbb{R}
\right) $\textit{\ via those in }$\mathfrak{su}\left( 4\right) $

We finish this section with an example which illustrates the utility of
passing to  $su\left( 4\right) $ for calculating exponentials in $%
\mathfrak{so}\left( 6,\text{ }%
\mathbb{R}
\right) $.

\begin{example}
\label{striking}Consider the matrix $X=\beta
(e_{4}e_{6}^{T}-e_{6}e_{4}^{T})+\delta (e_{6}e_{1}^{T}-e_{1}e_{6}^{T})$, for
some $\beta ,$ $\delta \in 
\mathbb{R}
$. Let us call the two summands $X_{1},$ $X_{2}$.

The summands $X_{1}$ and $X_{2}$ do\underline{ not anticommute or commute},
as can be easily verified. While the individual exponentials of $X_{1}$ and $%
X_{2}$ are easily found (both have cubic minimal polynomials), their sum,
without availing of the isomorphism with $\mathfrak{su}\left( 4\right) $,
presents a greater challenge. In fact, $X$ has a quintic minimal polynomial
as a brute force calculation, which we eschew, shows. On the other hand, $%
\Psi _{6}^{-1}(X)$ has a quadratic minimal polynomial!

Computing $W=\Psi _{6}^{-1}(X)\in \mathfrak{su}\left( 4\right) $, we find
that it is $\frac{i\beta }{2}\sigma _{x}\otimes \sigma _{x}+\frac{i(\gamma
-\alpha )}{2}\sigma _{z}\otimes \sigma _{x}=Z_{1}+Z_{2}$. In keeping with
the fact that $\Psi _{6}$ is a Lie algebra isomorphism, we see that $%
[Z_{1},Z_{2}]\neq 0$. However, $Z_{1}Z_{2}=-Z_{2}Z_{1}$. Thus, $W$'s minimal
poynomial is quadratic and one finds 
\begin{equation*}
e^{W}=cI_{4}+(\frac{s}{\lambda })[\frac{i\beta }{2}\sigma _{x}\otimes \sigma
_{x}+\frac{i(\gamma -\alpha )}{2}\sigma _{z}\otimes \sigma _{x}]
\end{equation*}%
where $\lambda =\frac{1}{2}\sqrt{\beta ^{2}+(\gamma -\alpha )^{2}}$, and $%
c=\cos (\lambda ),$ $s=\sin (\lambda )$. We next find $\Lambda =\theta _{%
\mathbb{C}
}(e^{W})$. It is given by 
\begin{equation*}
\Lambda =cI_{8}+\frac{s}{\lambda }[\frac{\beta }{2}(\sigma _{x}\otimes
\sigma _{x}\otimes i\sigma _{y})+\frac{(\gamma -\alpha )}{2}(\sigma
_{z}\otimes \otimes \sigma _{x}\otimes i\sigma _{y})]
\end{equation*}

Hence 
\begin{equation*}
\Lambda ^{T}=cI_{8}-\frac{s}{\lambda }[\frac{\beta }{2}(\sigma _{x}\otimes
\sigma _{x}\otimes i\sigma _{y})+\frac{(\gamma -\alpha )}{2}(\sigma
_{z}\otimes \otimes \sigma _{x}\otimes i\sigma _{y})]
\end{equation*}

To find $e^{X}$, we compute $\Lambda Y_{i}\Lambda ^{T},$ $i=1,\ldots ,6$.
Suppose $\Lambda Y_{j}\Lambda ^{T}=\sum_{i=1}^{6}c_{ij}Y_{i}$, then $%
e^{X}=(c_{ij})$.

To that end, we need the following:

\begin{itemize}
\item $\Lambda Y_{1}\Lambda ^{T}=\Lambda (\sigma _{x}\otimes (i\sigma
_{y})\otimes (-\sigma _{z}))\Lambda ^{T}$ is given by 
\begin{equation*}
\lbrack c^{2}-\frac{s^{2}}{4\lambda ^{2}}((\gamma -\alpha )^{2}-\beta
^{2})]Y_{1}+\frac{2s^{2}\beta (\gamma -\alpha )}{4\lambda ^{2}}Y_{4}+\frac{%
cs(\gamma -\alpha )}{\lambda }Y_{6}
\end{equation*}

\item $\Lambda Y_{2}\Lambda ^{T}=\Lambda (i\sigma _{y}\otimes \sigma
_{x}\otimes \sigma _{x})\Lambda ^{T}$ is given by%
\begin{equation*}
\frac{\left( \beta ^{2}+\left( \gamma -\alpha \right) ^{2}\right)
c^{2}+s^{2}\beta ^{2}+s^{2}\left( \gamma -\alpha \right) ^{2}}{\beta
^{2}+\left( \gamma -\alpha \right) ^{2}}Y_{3}=Y_{2}
\end{equation*}

\item $\Lambda Y_{3}\Lambda ^{T}=\Lambda (i\sigma _{y}\otimes I_{2}\otimes
\sigma _{z})\Lambda ^{T}$ is given by 
\begin{equation*}
(c^{2}+\frac{s^{2}[\beta ^{2}+(\gamma -\alpha )^{2}]}{4\lambda ^{2}}%
Y_{3}=Y_{3}
\end{equation*}

\item $\Lambda Y_{4}\Lambda ^{T}=\Lambda (-\sigma _{z}\otimes (i\sigma
_{y})\otimes \sigma _{z})\Lambda ^{T}$ is given by 
\begin{equation*}
\frac{2s^{2}\beta (\gamma -\alpha )}{\beta ^{2}+(\gamma -\alpha )^{2}}Y_{1}+%
\frac{(\gamma -\alpha )^{2}+\beta ^{2}(c^{2}-s^{2})}{\beta ^{2}+(\gamma
-\alpha )^{2}}Y_{4}-\frac{2cs\beta }{2\lambda }Y_{6}
\end{equation*}

\item $\Lambda Y_{5}\Lambda ^{T}=\Lambda (-I_{2}\otimes (i\sigma
_{y})\otimes \sigma _{x})\Lambda ^{T}$ is given by 
\begin{equation*}
\frac{\beta ^{2}+(\gamma -\alpha )^{2}(c^{2}+s^{2})}{\beta ^{2}+(\gamma
-\alpha )^{2}}Y_{5}=Y_{5}
\end{equation*}

\item $\Lambda Y_{6}\Lambda ^{T}=\Lambda ((-i\sigma _{y})\otimes \sigma
_{z}\otimes \sigma _{x})\Lambda ^{T}$ is given by 
\begin{equation*}
-\frac{2cs(\gamma -\alpha )}{2\lambda }Y_{1}+\frac{cs\beta }{\lambda }%
Y_{4}+(c^{2}-s^{2})Y_{6}
\end{equation*}
\end{itemize}

Hence, 
\begin{equation}
\exp (X)=\left( 
\begin{array}{cccccc}
\frac{\beta ^{2}+(\gamma -\alpha )^{2}(c^{2}-s^{2})}{\beta ^{2}+(\gamma
-\alpha )^{2}} & 0 & 0 & \frac{2s^{2}\beta (\gamma -\alpha )}{\beta
^{2}+(\gamma -\alpha )^{2}} & 0 & -\frac{cs(\gamma -\alpha )}{\lambda } \\ 
0 & 1 & 0 & 0 & 0 & 0 \\ 
0 & 0 & 1 & 0 & 0 & 0 \\ 
\frac{2s^{2}\beta (\gamma -\alpha )}{\beta ^{2}+(\gamma -\alpha )^{2}} & 0 & 
0 & \frac{(\gamma -\alpha )^{2}+\beta ^{2}(c^{2}-s^{2})}{\beta ^{2}+(\gamma
-\alpha )^{2}} & 0 & \frac{cs\beta }{\lambda } \\ 
0 & 0 & 0 & 0 & 1 & 0 \\ 
\frac{cs(\gamma -\alpha )}{\lambda } & 0 & 0 & -\frac{cs\beta }{\lambda } & 0
& c^{2}-s^{2}%
\end{array}%
\right)
\end{equation}
\end{example}

\section{Minimal Polynomials of Matrices in $\mathfrak{su}\left( 4\right) $}

In this section the minimal polynomials of matrices $X\in \mathfrak{su}%
\left( 4\right) $, is characterized completely. Thus, the problem of
exponentiation in $\mathfrak{su}\left( 4\right) $ and hence in $\mathfrak{so}%
\left( 6,\text{ }%
\mathbb{R}
\right) $ admits solutions which are constructive. The characterization of
the minimal polynomials will involve verifiable conditions on the $E_{k}(X),$
$k=2,$ $3,$ $4$. Recall $E_{k}(X)$ is the sum of all $k\times k$ principal
minors of $X$ and these are easy to compute.

The initial observation, which follows from arguments similar to those in
Proposition \ref{CharMinPolyofHMatrices} and Proposition \ref{shortlist}, is
that the minimal polynomial, $m_{X}$, of $X\in \mathfrak{su}\left( 4\right) $%
, has the following property:

\begin{description}
\item[A)] If the degree of $m_{X}$ is even, then the coefficients of all the
even powers of $x$ in it are real, while those of the odd powers are purely
imaginary.

\item[B)] If the degree of $m_{X}$ is odd, then the coefficients of all the
odd powers of $x$ in it are real, while those of the even powers are purely
imaginary.
\end{description}

This observation can be honed into the following result:

\begin{theorem}
\label{MinPolysufour}Let $X$ be a non-zero matrix in $\mathfrak{su}\left(
4\right) $. Then the structure of the minimal polynomials of $X$ is given by

\begin{enumerate}
\item $X$ has the minimal polynomial $x^{2}+\lambda ^{2}$, with $\lambda \in 
\mathbb{R}
$ non-zero, iff $E_{3}=0,$ $E_{2}\neq 0$ and $E_{4}=\frac{1}{4}(E_{2})^{2}$.

\item $X$ has the minimal polynomial $x^{2}+i\gamma x+\lambda ^{2}$, with $%
\gamma ,\lambda \in 
\mathbb{R}
$ both non-zero iff $E_{2}>0,$ $E_{4}=-\frac{1}{12}(E_{2})^{2},$ $E_{3}=8i(%
\sqrt{\frac{E_{2}}{6}})^{3}$.

\item $X$ has the minimal polynomial $x^{3}+\theta ^{2}x$, with $\theta \in 
\mathbb{R}
$ non-zero, iff $E_{3}=0=E_{4}$ and $E_{2}>0$.

\item $X$ has minimal polynomial $x^{3}+i\gamma x^{2}+\theta ^{2}x$, with $%
\gamma ,$ $\theta \in 
\mathbb{R}
$ both non-zero iff $E_{2}>0$ and $E_{3}$ is either $+2i(\sqrt{\frac{E_{2}}{3%
}})^{3}$ or $-2i(\sqrt{\frac{E_{2}}{3}})^{3}$.

\item $X$ has minimal polynomial $x^{3}+i\gamma x^{2}+\theta ^{2}x+i\delta $%
, with $\gamma ,$ $\theta ,$ $\delta \in 
\mathbb{R}
$, all non-zero iff $E_{4}\neq 0$, and 
\begin{equation}
16E_{2}^{4}E_{4}-4E_{2}^{3}E_{3}^{2}-128E_{2}^{2}E_{4}^{2}+144E_{2}E_{3}^{2}E_{4}-27E_{3}^{4}+256E_{4}^{3}=0
\label{ResVanishes}
\end{equation}%
and at least one of the conditions in each of items $1)$ and $2)$ above is
violated.

\item The minimal polynomial of $X$ is its characterisitic polynomial iff
the condition in Equation \ref{ResVanishes} is violated.
\end{enumerate}

Furthermore, in each of these cases the coefficients of the minimal
polynomial can be determined \underline{constructively} from the $E_{k}$.
\end{theorem}

\noindent \textbf{Proof:} First, since $X$ is skew-Hermitian, so is every
principal submatrix of $X$. Since the determinant of an even sized (resp.
odd sized) skew-Hermitian matrix is real (resp. purely imaginary) it follows
that $E_{2},E_{4}\in 
\mathbb{R}
$ and $iE_{3}\in 
\mathbb{R}
$.

Next, since $X$ is diagonalizable its minimal polynomial has distinct roots.
In view of $E_{1}=0$ and $X\neq 0$, the following are the root
configurations of the characteristic polynomial, $p_{X}(x)$, which lead to
its minimal polynomial, $m_{X}(x)$, being of strictly lower degree than $4$:

\begin{description}
\item[Case 1)] The two distinct roots of $p_{X}$ are $ia$ and $-ia$, each
with multiplicity $2$, and $a\in 
\mathbb{R}
$ non-zero. In this case $m_{X}=x^{2}+a^{2}$.

\item[Case 2)] The two distinct roots of $p_{X}$ are $ia$ and $ib$ with $i)$ 
$a,b$ non-zero real; and $ii)$ the former repeated thrice and the latter
once. In this case, necessarily $b=-3a$. In this case $%
m_{X}=x^{2}+2iax+3a^{2}$.

\item[Case 3)] The three distinct roots of $p_{X}$ are $0$ (repeated twice)
and $ia$ and $-ia$ of multiplicity one each (with $a\in 
\mathbb{R}
$ non-zero). In this case $m_{X}=x^{3}+a^{2}x$.

\item[Case 4)] The three distinct roots of $p_{X}$ are $ia$, $ib$ and $0$,
with first repeated twice and the latter two of multiplicity one each. Once
again $a,b\in 
\mathbb{R}
$ are non-zero. In this case, necessarily $b=-2a$ and $%
m_{X}=x^{3}+iax^{2}+2a^{2}x$.

\item[Case 5)] The three distinct roots of $p_{X}$ are $ia$, $ib$ and $ic$,
with $i)$ $a,b,c\in 
\mathbb{R}
$ and $abc\neq 0$; and $ii)$ the multiplicity of $ia$ is two, while that of
the other roots is one each. In this case necessarily, $b+c=-2a$.
Furthermore, 
\begin{equation*}
m_{X}=(x-ia)(x-ib)(x-ic)=x^{3}-i(a+b+c)x^{2}+(-a^{2}-ab-ac)x+iabc
\end{equation*}%
Since, $b+c=-2a$, this simplifies, for the moment, to%
\begin{equation*}
m_{x}=x^{3}+iax^{2}+a^{2}x+iabc
\end{equation*}

\item[Case 6)] All roots of $p_{X}(x)$ are distinct. In this case the
minimal polynomial is $p_{X}$.
\end{description}

To now characterize these root configurations, without having to find the
roots, we note that $E_{k}=S_{k},$ $\forall k$, where $S_{k}$ is, of course,
the $k$th elementary symmetric polynomial of the roots of the characteristic
polynomial. So we have

\begin{description}
\item[Case 1)] In this case $%
E_{2}=-a^{2}+a^{2}+a^{2}+a^{2}+a^{2}-a^{2}=2a^{2}$. Similarly $E_{3}=0$ and $%
E_{4}=a^{4}$. So for $X$ to have the minimal polynomial $x^{2}+\lambda ^{2}$%
, it is necessary that $E_{3}=0,$ $E_{2}>0$ and $E_{4}=\frac{1}{4}(E_{2})^{2}
$. Furthermore, $\lambda =\sqrt{\frac{E_{2}}{2}}$.

\item The converse is also true. If these conditions on the $E_{k}$ hold, 
\begin{equation*}
p_{X}=x^{4}+E_{2}x^{2}+E_{4}=x^{4}+E_{2}x^{2}+(\frac{E_{2}}{2})^{2}
\end{equation*}%
Quite clearly this is a quadratic in $x^{2}$, leading to the eigenvalues
being of the form $ia$ and $-ia$, each repeated twice, with $a$ the positive
square root of $\frac{E_{2}}{2}$, which, of course leads to $m_{X}=x^{2}+%
\frac{E_{2}}{2}$.

\item[Case 2)] In this case $E_{2}=6a^{2}$, while $E_{3}=8ia^{3}$ and
finally, $E_{4}=-3a^{4}$. From this it follows that a necessary condition
for $X$ to have the minimal polynomial 
\begin{equation*}
m_{X}(x)=x^{2}+i\gamma x+c^{2}
\end{equation*}%
is that $E_{2}>0,$ $E_{3}=8i[\frac{E_{2}}{6}]^{\frac{3}{2}}$ and $E_{4}=-%
\frac{1}{12}(E_{2})^{2}$.

\item The converse also holds. Indeed, in this case, $p_{X}$ has a triple
root. Hence $p_{X}^{^{\prime }}$ has a double root and this double root is
one of the roots of $p_{X}^{^{\prime \prime }}$. Now 
\begin{equation*}
p_{X}^{^{\prime \prime }}=12x^{2}+2E_{2}
\end{equation*}%
Its roots are $i\sqrt{\frac{E_{2}}{6}}$ and $-i\sqrt{\frac{E_{2}}{6}}$. Only
one of these can be a root of $p_{X}$, since neither is $-3$ times the other
and $p_{X}$ has only one multiple root. We calculate 
\begin{equation*}
p_{X}(i\sqrt{\frac{E_{2}}{6}})=\frac{E_{2}^{2}}{36}-\frac{E_{2}^{2}}{6}+%
\frac{8E_{2}^{2}}{36}-\frac{E_{2}^{2}}{12}=0
\end{equation*}%
Here we have made use of the necessary conditions $E_{2}>0,$ $E_{3}=8i(\sqrt{%
\frac{E_{2}}{6})^{3}}$ and $E_{4}=-\frac{1}{12}(E_{2})^{2}$.

\item Thus, sufficiency has also been verified. Finally, note that the
coefficients of the minimal polynomial satisfy $\gamma =2a,c^{2}=3a^{2}$.
Both can be obtained \textbf{without} finding $a$. Clearly, $c^{2}=\frac{%
E_{2}}{2}$ and to find $\gamma $ we look at the sign of the purely imaginary
number $E_{3}$. Its sign coincides with the sign of $\gamma $, and the
actual value of $\gamma $ is then found from, say, just $E_{2}$.

\item[Case 3)] In this case, we find $E_{2}=a^{2}$ and that $E_{3}=0=E_{4}$.
So the stated conditions are obviously necessary. They are also sufficient,
since under these conditions the characteristic polynomial is 
\begin{equation*}
p_{X}(x)=x^{4}+E_{2}x^{2}=x^{2}(x^{2}+E_{2})
\end{equation*}%
Since $E_{2}>0$, its roots are obviously $0$ (repeated twice) and $i\sqrt{%
E_{2}}$ and $-i\sqrt{E_{2}}$.

\item Finally, the minimal polynomial, in this case, is $m_{X}=x^{3}+c^{2}x$%
, and $c^{2}$ is evidently uniquely determined as $c^{2}=E_{2}$.

\item[Case 4)] In this case $E_{2}=3a^{2},$ $E_{3}=2ia^{3},$ $E_{4}=0$. So
necessarily $E_{2}>0$ and $E_{3}$ is plus or minus $2i(\frac{E_{2}}{3})^{%
\frac{3}{2}}$ and $E_{4}=0$.

\item To verify the converse note that, if the stated conditions on $E_{2},$ 
$E_{3},$ $E_{4}$ hold then 
\begin{equation*}
p_{X}(x)=x^{4}+E_{2}x^{2}-2i(\frac{E_{2}}{3})^{\frac{3}{2}%
}x=x(x^{3}+E_{2}x-2i(\frac{E_{2}}{3})^{\frac{3}{2}})
\end{equation*}%
So $0$ is a single root and the remaining roots of $p_{X}$ are the roots of 
\begin{equation*}
q(x)=x^{3}+E_{2}x-2i(\frac{E_{2}}{3})^{\frac{3}{2}}
\end{equation*}%
To show that $q(x)$, and thus $p_{X}$, has a double root we compute 
\begin{equation*}
q^{^{\prime }}(x)=3x^{2}+E_{2}
\end{equation*}%
Its roots are $x=i\sqrt{\frac{E_{2}}{3}}$ and $x=-i\sqrt{\frac{E_{2}}{3}}$.
We check if one of these roots is a root of $p_{X}$. We find, if $E_{3}=2i(%
\frac{E_{2}}{3})^{\frac{3}{2}}$, then 
\begin{equation*}
p(i\sqrt{\frac{E_{2}}{3}})=0
\end{equation*}%
If $E_{3}=-2i(\frac{E_{2}}{3})^{\frac{3}{2}}$, then 
\begin{equation*}
p(-i\sqrt{\frac{E_{2}}{3}})=0
\end{equation*}%
So indeed the stated conditions are sufficient as well.

\item Finally, to determine the coefficients of $m_{X}(x)=x^{3}+i\gamma
x^{2}+\theta ^{2}x$, we note that since $m_{X}$ is also $%
x^{3}+iax^{2}+2a^{2}x$, we must have $\theta ^{2}=2a^{2}=\frac{2}{3}E_{2}$,
and that $\gamma $ is plus or minus $i\sqrt{\frac{E_{2}}{3}}$, depending on
the sign of the non-zero purely imaginary number $E_{3}$.

\item[Case 5)] $X$ has a minimal polynomial, which is of lower degree than $4
$, iff $p_{X}$ has a repeated root. Now $p_{X}$ has a repeated root iff it
and its derivative have a common root. The latter condition obtains iff the
resultant of $p_{X}$ and $p_{X}^{^{\prime }}$ vanish. This condition is
precisely the validity of Equation (\ref{ResVanishes}). The remaining
conditions ensure that this repeated root configuration is not one of the
preceding cases, and thus has to correspond to the root configuration $\{ia,$
$ia,$ $ib,$ $ic\}$, with $abc\neq 0$.

\item To determine the coefficients of $m_{X}$, we first note that, since $%
c=-(b+2a)$ that 
\begin{equation*}
m_{X}=x^{3}+(ia)x^{2}+(2a^{2}-bc)x+iabc
\end{equation*}%
Let us write this 
\begin{equation*}
m_{X}=x^{3}+c_{1}x^{2}+c_{2}x+c_{3}
\end{equation*}%
Now, $E_{2}=-a^{2}-2a(b+c)-bc=3a^{2}-bc$. Thus, $c_{2}=E_{2}-a^{2}$.
Similarly, $c_{3}=i\frac{E_{4}}{a}$. Hence, 
\begin{equation*}
m_{X}=x^{3}+(ia)x^{2}+(E_{2}-a)x+i\frac{E_{4}}{a}
\end{equation*}%
So to fully find $m_{X}$ we need $a$. There are two ways to proceed, the
second of which is relegated to Remark \ref{NewTechnique} below. The first
method proceeds as follows. Note first that 
\begin{equation*}
E_{3}=i(2a^{3}-2abc)
\end{equation*}%
Since $E_{2}=3a^{2}-bc$, we find 
\begin{equation*}
E_{3}=i[-4a^{3}+(2a)E_{2}]
\end{equation*}%
Equivalently, $iE_{3}=4a^{3}-(2a)E_{2}$. Hence, $a$ is a root of the cubic 
\begin{equation}
c(x)=4X^{3}-(2E_{2})x-iE_{3}=0  \label{cubicroot}
\end{equation}%
Since $E_{2}$ and $iE_{3}$ are real, this cubic has at least one real root.
If this cubic has only one real root then, that real root gives $a$ and we
are done. If it has three real roots, say $\alpha ,\beta ,\gamma $, then by
construction precisely one of $\{i\alpha ,$ $i\beta ,$ $i\gamma \}$ is a
double root of $p_{X}$. So we evaluate $p_{X}$ and $p_{X}^{^{\prime }}$ at
these points and see at which of these both vanish. That gives $a$ and hence 
$m_{X}$.$\diamondsuit $
\end{description}

\begin{remark}
\label{NewTechnique}A second method to determine the coefficients of the
minimal polynomial, $m_{X}(x)$, in Case $5$ of the previous theorem, is
now discussed. This method requires only the solution of a quadratic
equation and works with $E_{4}$ and $\left\Vert X\right\Vert _{F}^{2}$.
Begin by observing that, since $X$ is a normal matrix it follows that 
\begin{equation*}
\left\Vert X\right\Vert _{F}^{2}=\left\vert ia\right\vert ^{2}+\left\vert
ia\right\vert ^{2}+\left\vert ib\right\vert ^{2}+\left\vert ic\right\vert
^{2}=2a^{2}+b^{2}+c^{2}
\end{equation*}

Now using $i)$ $2a^{2}+b^{2}+c^{2}=2a^{2}+(b+c)^{2}-2bc$ and $ii)$ $b+c=-2a$%
, we find that 
\begin{equation*}
\left\Vert X\right\Vert _{F}^{2}=6a^{2}-\frac{2a^{2}bc}{a^{2}}=6a^{2}-\frac{%
2E_{4}}{a^{2}}
\end{equation*}%
Hence $a^{2}$ is a solution of the quadratic 
\begin{equation*}
6x^{2}-\left\Vert X\right\Vert _{F}^{2}x-2E_{4}=0
\end{equation*}%
By construction, this quadratic has at least one positive real solution
(and, thus, in fact, both solutions must be real). Thus, this gives upto
four choices of $a$. The correct one is that value which yields $%
iE_{3}=4a^{3}-2aE_{2}$.
\end{remark}

\begin{remark}
$e^{X}$ can be found for any $X\in \mathfrak{su}\left( 4\right) $ satisfying
the first 3 cases of Theorem \ref{MinPolysufour} by using the formulae
presented in Theorem \ref{ExpFromMinPolyList}. For cases $4)$ and $5)$ of
Theorem \ref{MinPolysufour} one can use Lagrange interpolation, i.e., $e^{X}$
is that polynomial in $X$ which takes on the value $e^{ir}$ at a root $%
ir,r\in 
\mathbb{R}
$ of the corresponding minimal polynomial. Note that the proof of Theorem %
\ref{MinPolysufour} supplies, as a byproduct, recipes to find the roots of
the minimal polynomial in cases $4)$ and $5)$. For case $6)$, if $E_{3}=0$,
then one can invoke case $IV)$ of Theorem \ref{ExpFromMinPolyList}.
Similarly, in Case $6)$ if $E_{4}=\det (X)=0$, then one can easily find the
roots of the characteristic polynomial. They are given by $0,$ $i\alpha ,$ $%
i\beta ,$ $-i(\alpha +\beta )$, with $\alpha \beta \neq 0$ and $\alpha \neq
\beta $ and $\alpha \neq -\beta $. These can be found by solving a cubic.
Finally, in Case $6)$, if neither $E_{3}$ nor $E_{4}$ is zero, then one has
to solve a quartic to find the eigenvalues, which, albeit, complicated, can
be found in closed form. One can then use Lagrange interpolation to find $%
e^{X}$. At any rate, as mentioned before, in the cases not susceptible to
the formulae in Theorem \ref{ExpFromMinPolyList}, it is of utility to first
investigate whether $X$ can be expressed as a sum of commuting summands,
each of which has a lower degree minimal polynomial. This is the case, for
instance, if either $X$ is purely imaginary or purely real, \cite{expisufour}.
\end{remark}

\section{${\mbox Spin}\left( 5\right) $ Reconsidered}

Section 3 started with a basis of $1$-vectors for ${\mbox C}l\left( 3,\text{ 
}0\right) $ (namely the Pauli basis) and applied the natural constructions
in Sec $2.3$ to arrive at a basis of $1$-vectors for ${\mbox C}l\left( 0,%
\text{ }5\right) $. The ability to produce a basis of $1$-vectors for ${%
\mbox C}l\left( 0,\text{ }6\right) $, starting from ${\mbox C}l\left( 0,%
\text{ }0\right) $, which lead to to $\tilde{J}_{8}$ playing a role in
reversion, naturally raises the question whether following that set of
iterative constructions could lead to something similar for ${\mbox C}%
l\left( 0,\text{ }5\right) $. We show below that this is the case and 
\underline{more importantly} that a slight variation of this construction
reveals a role in reversion for yet another matrix in the $\mathbb{H}\otimes 
\mathbb{H}$ basis for $M\left( 4,\text{ }%
\mathbb{R}
\right) $, viz., the matrix $M_{j\otimes 1}$! In the process, a natural
interpretation of the matrix $X\left( z_{0},\text{ }z_{1},\text{ }%
z_{2}\right) $ of Remark \ref{complicated} is also found.

Let us first show how $\widetilde{J}_{4}$ arises. We start with ${\mbox C}%
l\left( 0,\text{ }1\right) $ and apply the construction \textbf{IC1} of Sec $%
2.3$ twice to arrive at a basis of $1$-vectors for ${\mbox C}l\left( 2,\text{
}3\right) $. Next we use \textbf{IC3} of Sec $2.3$ to arrive at a basis of $1
$-vectors for ${\mbox C}l\left( 4,\text{ }1\right) $, and then finally use 
\textbf{IC2} of Sec $2.3$ to arrive at a basis of $1$-vectors for ${\mbox C}%
l\left( 0,\text{ }5\right) $.

We begin with $\{i\}$ as the obvious basis of $1$-vectors for ${\mbox C}%
l\left( 0,\text{ }1\right) $. This gives $\{\sigma _{x},$ $i\sigma _{y},$ $%
i\sigma _{z}\}$ as a basis for ${\mbox C}l\left( 1,\text{ }2\right) $. This
then yields the following five matrices as a basis of $1$-vectors for ${%
\mbox C}l\left( 2,\text{ }3\right) $: 
\begin{equation*}
\left( 
\begin{array}{cc}
\sigma _{x} & 0 \\ 
0 & -\sigma _{x}%
\end{array}%
\right) ;\text{ }\left( 
\begin{array}{cc}
0 & I_{2} \\ 
I_{2} & 0%
\end{array}%
\right) ;\text{ }\left( 
\begin{array}{cc}
i\sigma _{y} & 0 \\ 
0 & -i\sigma _{y}%
\end{array}%
\right) ;\text{ }\left( 
\begin{array}{cc}
i\sigma _{z} & 0 \\ 
0 & -i\sigma _{z}%
\end{array}%
\right) ;\text{ }\left( 
\begin{array}{cc}
0 & I_{2} \\ 
-I_{2} & 0%
\end{array}%
\right) 
\end{equation*}

Written more succintly this last basis is 
\begin{equation*}
\{\sigma _{z}\otimes \sigma _{x},\text{ }\sigma _{x}\otimes I_{2},\text{ }%
\sigma _{z}\otimes i\sigma _{y},\text{ }\sigma _{z}\otimes i\sigma _{z},%
\text{ }i\sigma _{y}\otimes I_{2}\}
\end{equation*}

We now find the basis of $1$-vectors for ${\mbox C}l\left( 4,\text{ }%
1\right) $ by applying \textbf{IC3}. This yields the following basis

\begin{center}
\begin{tabular}{lllll}
$\tilde{e}_{1}$ & $=$ & $\sigma _{z}\otimes \sigma _{x}$ & $=$ & $\sigma
_{z}\otimes \sigma _{x}$ \\ 
$\tilde{e}_{2}$ & $=$ & $(\sigma _{x}\otimes I_{2})(\sigma _{z}\otimes
\sigma _{x})$ & $=$ & $-i\sigma _{y}\otimes \sigma _{x}$ \\ 
$\tilde{e}_{3}$ & $=$ & $(\sigma _{z}\otimes i\sigma _{y})(\sigma
_{z}\otimes \sigma _{x})$ & $=$ & $I_{2}\otimes \sigma _{z}$ \\ 
$\tilde{e}_{4}$ & $=$ & $(\sigma _{z}\otimes i\sigma _{z})(\sigma
_{z}\otimes \sigma _{x})$ & $=$ & $I_{2}\otimes (-\sigma _{z})$ \\ 
$\tilde{e}_{5}$ & $=$ & $(i\sigma _{y}\otimes I_{2})(\sigma _{z}\otimes
\sigma _{x})$ & $=$ & $-\sigma _{x}\otimes -\sigma _{x}$%
\end{tabular}
\end{center}

Relabelling this last basis to be consistent with signature to obtain the
basis of $1$-vectors for ${\mbox C}l\left( 4,\text{ }1\right) $ yields 
\begin{equation*}
\{h_{1}=\sigma _{z}\otimes \sigma _{x},\text{ }h_{2}=I_{2}\otimes \sigma
_{z},\text{ }h_{3}=I_{2}\otimes -\sigma _{y},\text{ }h_{4}=-\sigma
_{x}\otimes \sigma _{x},\text{ }h_{5}=-i\sigma _{y}\otimes \sigma _{x}\}
\end{equation*}

Finally applying \textbf{IC2} to this last basis gives a basis of $1$%
-vectors for ${\mbox C}l\left( 0,\text{ }5\right) $. To that end we first
find 
\begin{equation*}
h_{1}h_{2}h_{3}h_{4}=-i\sigma _{y}\otimes i\sigma _{x}
\end{equation*}

This then yields the desired basis of $1$-vectors for ${\mbox C}l\left( 0,%
\text{ }5\right) $ as follows:

\begin{center}
\begin{tabular}{lllll}
$f_{1}$ & $=$ & $h_{1}(-i\sigma _{y}\otimes i\sigma _{x})$ & $=$ & $-\sigma
_{x}\otimes (iI_{2})$ \\ 
$f_{2}$ & $=$ & $h_{2}(-i\sigma _{y}\otimes i\sigma _{x})$ & $=$ & $i\sigma
_{y}\otimes \sigma _{y}$ \\ 
$f_{3}$ & $=$ & $h_{3}(-i\sigma _{y}\otimes i\sigma _{x})$ & $=$ & $i\sigma
_{y}\otimes \sigma _{z}$ \\ 
$f_{4}$ & $=$ & $h_{4}(-i\sigma _{y}\otimes i\sigma _{x})$ & $=$ & $-\sigma
_{z}\otimes (iI_{2})$ \\ 
$f_{5}$ & $=$ & $h_{5}$ & $=$ & $-i\sigma _{y}\otimes \sigma _{x}$%
\end{tabular}
\end{center}

Evidently, we may replace those $f_{i}$'s with a negative sign by their
negatives without losing any virtues. Let us relabel this basis as $%
\{g_{k}\mid k=1,\ldots ,5\}$.

\begin{proposition}
\label{IIBasisforCliff5}With respect to the basis of $1$-vectors for ${\mbox
C}l\left( 0,\text{ }5\right) $ given by the matrices: 
\begin{equation*}
\{g_{1} = \sigma _{x}\otimes I_{2},\text{ }g_{2}=i\sigma _{y}\otimes \sigma
_{y},\text{ }g_{3}=i\sigma _{y}\otimes \sigma _{z},\text{ }g_{4}=i\sigma
_{z}\otimes I_{2},\text{ }g_{5}=i\sigma _{y}\otimes \sigma _{x}\}
\end{equation*}%
reversion on ${\mbox C}l\left( 0,\text{ }5\right) =M\left( 4,\text{ }%
\mathbb{C}
\right) $ is described by 
\begin{equation*}
\Phi ^{rev}(X)=\widetilde{J}_{4}^{-1}X^{T}\widetilde{J}_{4}
\end{equation*}
\end{proposition}

\noindent \textbf{Proof:} It suffices to verfiy that $\widetilde{J}%
_{4}^{-1}g_{i}^{T}\widetilde{J}_{4}=g_{i},$ $\forall i=1,\ldots ,5$ We
verify this only for $g_{1}$ by way of illustration 
\begin{equation*}
\tilde{J}_{4}^{-1}g_{1}^{T}\tilde{J}_{4}=(I_{2}\otimes -i\sigma _{y})(\sigma
_{x}\otimes iI_{2})(I_{2}\otimes i\sigma _{y})=\sigma _{x}\otimes
iI_{2}=g_{1}
\end{equation*}%
$\diamondsuit $

\begin{remark}
A quick calculation shows that $J_{4}^{-1}g_{1}^{T}J_{4}$ is actually $%
-g_{1} $. Thus, reversion, for this basis cannot involve $J_{4}$.
\end{remark}

\begin{remark}
\label{complicated2}Let us examine what a typical $1$-vector looks like,
with respect to the basis $\{g_{k}\}$ in Proposition \ref{IIBasisforCliff5}.
It is given by the following matrix 
\begin{equation*}
\left( 
\begin{array}{cccc}
id & 0 & c+ia & e-bi \\ 
0 & id & e+ib & -c+ia \\ 
-c+ia & -e+ib & -id & 0 \\ 
-e-ib & c+ia & 0 & -id%
\end{array}%
\right) 
\end{equation*}

(with $a,$ $b,$ $c,$ $d,$ $e\in 
\mathbb{R}
$) But this matrix is precisely $X\left( z_{0},\text{ }z_{i},\text{ }%
z_{2}\right) $ described in Remark \ref{complicated}, with $%
z_{0}=c+ia,z_{1}=e+ib,z_{2}=id$. This gives a different motivation for this
matrix in \cite{portei}. Notice that $z_{2}$ being allowed to be possibly
not purely imaginary is precisely the obstruction to $X\left( z_{0},\text{ }%
z_{i},\text{ }z_{2}\right) $ to being anti-Hermitian.It should be pointed
out that the basis $\{g_{i}\mid i=1,\ldots ,5\}$ of Proposition \ref%
{IIBasisforCliff5} is not present in \cite{portei}, since for identification
of $\mbox Spin\left( 5\right) $, \cite{portei} works in ${\mbox C}l\left( 0,%
\text{ }4\right) $.
\end{remark}

We now discuss a slight variation on this construction. Everything remains
verbatim upto the basis of $1$-vectors for ${\mbox C}l\left( 2,\text{ }%
3\right) $. However, for the production of a basis of $1$-vectors for ${%
\mbox C}l\left( 4,\text{ }1\right) $ we proceed alternatively in the
following manner:

\begin{center}
\begin{tabular}{lllll}
$\hat{e}_{1}$ & $=$ & $\sigma _{x}\otimes I_{2}$ & $=$ & $\sigma _{x}\otimes
I_{2}$ \\ 
$\hat{e}_{2}$ & $=$ & $(\sigma _{z}\otimes \sigma _{x})(\sigma _{x}\otimes
I_{2})$ & $=$ & $i\sigma _{y}\otimes \sigma _{x}$ \\ 
$\hat{e}_{3}$ & $=$ & $(\sigma _{z}\otimes i\sigma _{y})(\sigma _{x}\otimes
I_{2})$ & $=$ & $-\sigma _{y}\otimes \sigma _{y}$ \\ 
$\hat{e}_{4}$ & $=$ & $(\sigma _{z}\otimes i\sigma _{z})(\sigma _{x}\otimes
I_{2})$ & $=$ & $-\sigma _{y}\otimes \sigma _{z}$ \\ 
$\hat{e}_{5}$ & $=$ & $(i\sigma _{y}\otimes I_{2})(\sigma _{x}\otimes I_{2})$
& $=$ & $\sigma _{z}\otimes I_{2}$%
\end{tabular}
\end{center}

In other words, we have interchanged the roles of $\sigma _{z}\otimes \sigma
_{x}$ and $\sigma _{x}\otimes I_{2}$ - the two $1$-vectors in ${\mbox C}%
l\left( 2,\text{ }3\right) $ which square to $+1$, cf., Remark \ref%
{Choiceofedoesnotmatter}.

Once again relabelling this basis to reflect signature, yields a basis of $1$%
-vectors for ${\mbox C}l\left( 4,\text{ }1\right) $ in the form 
\begin{equation*}
\hat{h}_{1}=\sigma _{x}\otimes I_{2},\text{ }\hat{h}_{2}=-\sigma _{y}\otimes
\sigma _{y},\text{ }\hat{h}_{3}=-\sigma _{y}\otimes \sigma _{z},\text{ }\hat{%
h}_{4}=\sigma _{z}\otimes I_{2},\text{ }\hat{h}_{5}=i\sigma _{y}\otimes
\sigma _{x}
\end{equation*}

Now applying \textbf{IC2} of Sec $2.3$, as before, we first calculate 
\begin{equation*}
\hat{h}_{1}\hat{h}_{2}\hat{h}_{3}\hat{h}_{4}=\sigma _{y}\otimes \sigma _{x}
\end{equation*}

This then yields yet another basis of $1$-vectors for ${\mbox C}l\left( 0,%
\text{ }5\right) $ given by 
\begin{eqnarray*}
\hat{f}_{1} &=&i\sigma _{z}\otimes \sigma _{x} \\
\hat{f}_{2} &=&I_{2}\otimes i\sigma _{z} \\
\hat{f}_{3} &=&-I_{2}\otimes i\sigma _{y} \\
\hat{f}_{4} &=&-i\sigma _{x}\otimes \sigma _{x} \\
\hat{f}_{5} &=&i\sigma _{y}\otimes \sigma _{x}
\end{eqnarray*}

We now ask what is the explicit form of reversion on ${\mbox C}l\left( 0,%
\text{ }5\right) $ for this basis of $1$-vectors. Once again the $\mathbb{H}%
\otimes \mathbb{H}$ basis for $M\left( 4,\text{ }%
\mathbb{R}
\right) $ comes to our aid to provide the following

\begin{theorem}
Let $\breve{J}_{4}=M_{j\otimes 1}$. Then reversion on ${\mbox C}l\left( 0,%
\text{ }5\right) $, with respect to the basis $\{\hat{f}_{i}\mid i=1,\ldots
,5\}$ of $1$-vectors, obtained above is 
\begin{equation*}
\Phi _{rev}(X)=\breve{J}_{4}^{-1}X^{T}\breve{J}_{4}
\end{equation*}
\end{theorem}

\noindent \textbf{Proof} The explicit form of $\breve{J}_{4}$ is 
\begin{equation*}
\breve{J}_{4} = \left ( 
\begin{array}{cccc}
0 & 0 & -1 & 0 \\ 
0 & 0 & 0 & 1 \\ 
1 & 0 & 0 & 0 \\ 
0 & -1 & 0 & 0%
\end{array}
\right )
\end{equation*}

It is more useful to write it as 
\begin{equation*}
\breve{J}_{4}=i\sigma _{y}\otimes -\sigma _{z}
\end{equation*}%
With this at hand it suffices, as usual, to confirm that 
\begin{equation*}
\breve{J}_{4}^{-1}\hat{f}_{i}^{T}\breve{J}_{4}=\hat{f}_{i},\text{ }\forall
i=1,\ldots ,5
\end{equation*}

We will content ourselves by displaying the calculations for $\hat{f}_{1}$.
Since $(i\sigma_{y})^{-1} = -i\sigma_{y}$ and $(-\sigma_{z})^{-1}
=-\sigma_{z}$ we obtain 
\begin{equation*}
\breve{J}_{4}^{-1}\hat{f}_{1}^{T}\breve{J}_{4} =
(i\sigma_{y}\otimes\sigma_{z})(i\sigma_{z}\otimes\sigma_{x})
(i\sigma_{y}\otimes -\sigma_{z}) = i\sigma_{z}\otimes \sigma_{x} = \hat{f}%
_{1}
\end{equation*}
$\diamondsuit$.

\section{Conclusions}

In this note we have derived explicit matrix realizations of the reversion
automorphism for ${\mbox C}l\left( 0,\text{ }5\right) $ and ${\mbox C}%
l\left( 0,\text{ }6\right) $, with respect to bases of $1$-vectors which are
natural from the point of view of the standard iterative procedures,
described in Section $2.3$. This also leads to a first principles approach
to the spin groups in these dimensions, in the sense that they are obtained
by working entirely in ${\mbox C}l\left( 0,\text{ }5\right) $ and ${\mbox C}%
l\left( 0,\text{ }6\right) $ respectively. These constructions are then used
to find closed form expressions for the exponentials of real antisymmetric
matrices of size $5\times 5$ and $6\times 6$. This is facilitated by the
derivation of explicit expressions for the minimal polynomials of matrices
in the Lie algebras of the corresponding spin groups. These expressions do
not require any spectral knowledge of the matrices in question. Two
important byproducts of this note are that it provides further evidence for
the importance of the isomorphism between $\mathbb{H}\otimes \mathbb{H}$ and 
$M\left( 4,\text{ }%
\mathbb{R}
\right) $, and what hopefully is a didactically appealing derivation of the
spin groups for $n=5,$ $6$.

There some questions whose study this work naturally suggests. We mention
two here:

\begin{itemize}
\item It would be useful to obtain expressions for minimal polynomials of
matrices in $\mathfrak{su}\left( 4\right) $ directly from their $\mathbb{H}%
\otimes \mathbb{H}$ representations, analogous to the formulae in \cite%
{minpolyi}. Specifically, if one writes an $X\in \mathfrak{su}\left(
4\right) $ as $Y+iZ$ with $Y,Z$ real matrices, then $Y^{T}=-Y$ and $Z^{T}=Z$%
. This is significant because any such work will also yield formulae for
minimal polynomials of the \underline{real} matrix $Y+Z$. Since such a
matrix is the most general traceless real $4\times 4$ matrix, the benefits
are obvious. In Section 6, while no knowledge of eigenvalues or eigenvectors
was needed, the diagonalizability of matrices in $\mathfrak{su}\left(
4\right) $ was heavily used. On the other hand, the methods in \cite%
{minpolyi} never used any such information. Since there are many important
non-diagonalizable matrices in $M\left( 4,\text{ }%
\mathbb{R}
\right) $, this would be of high utility.

\item It is important to be able to invert the covering maps $\Phi _{5}$ and 
$\Phi _{6}$. One application of this would be the ability to deduce
factorizations of matrices in $SO\left( n,\text{ }%
\mathbb{R}
\right) $, for $n=5,$ $6$, from those for matrices in their spin groups. The
inversion of these maps requires solving a system polynomial equations in
several variables which are essentially quadratic. For a satisfactory
solution to this problem, a first step would be useful parametrizations or
representations of elements in their spin groups. A first attempt at this is
provided in the appendix for $Sp\left( 4\right) $. This representation may
be of independent interest.
\end{itemize}

\section{Appendix - A Representation of $Sp\left( 4\right) $}

In this section we discuss a representation of an element of $Sp\left(
4\right) $, which is partially motivated by the question of inverting the
covering map of $SO\left( 5,\text{ }%
\mathbb{R}
\right) $, and may be of independent interest. The reason for choosing $%
Sp\left( 4\right) $ rather than its variants ($\widehat{Sp}\left( 4\right) $%
, for instance) is that just as those variants were more amenable for
certain purposes [such as computing determinants- see Remark \ref%
{BlockStructureofWideHatsp}], the block structure of $Sp\left( 4\right) $ is
easier to describe matrix theoretically.

Loosely speaking the main observation is that every element of $Sp\left(
4\right) $ is a $\theta _{\mathbb{H}}$ matrix $\left( 
\begin{array}{cc}
A & B \\ 
-\bar{B} & \bar{A}%
\end{array}%
\right) $ in which $A$ is a contraction, and $B$ is essentially determined
by a square root of $I-A^{\ast }A$, which generically differs from defect of 
$A$ by a diagonal factor. The defect of $A$ is defined to be the unique
positive square root of $I-A^{\ast }A$.

The representation provided is not quite a parametrization since it requires
12 parameters and not 10, as the dimension of $Sp\left( 4\right) $ would
suggest. This is primarily due to the invocation of the singular value
decomposition of $A$. Nevertheless we believe it is computationally
tractable.

Consider, therefore, $X\in Sp\left( 4\right) $. It equals $\theta _{\mathbb{H%
}}(Y)$ for some unitary element $Y\in M\left( 2,\text{ }\mathbb{H}\right) $.
Writing $Y=A+Bj$, with $A,$ $B\in M\left( 2,\text{ }%
\mathbb{C}
\right) $, $Y$'s unitarity is equivalent to the equations 
\begin{eqnarray*}
A^{\ast }A+(\bar{B})^{\ast }(\bar{B}) &=&I_{2} \\
A^{\ast }B &=&B^{T}\bar{A}
\end{eqnarray*}

The second condition is, of course, the same as saying that the matrix $%
A^{*}B$ is symmetric.

The first condition says that the matrix $A$ is a contraction and that the
matrix $\bar{B}$ is one possible square root of the positive semidefinite
matrix $I_{2}-A^{\ast }A$ (recall that a matrix $Q\in M\left( n,\text{ }%
\mathbb{C}
\right) $ is a square root of a positive semidefinite matrix $P$ if $Q^{\ast
}Q=P$).

In order to extract more information from this, first observe that $A$ being
a contraction is equivalent to its largest singular value being atmost one.
Thus, 
\begin{equation*}
A = U\left (%
\begin{array}{cc}
\sigma_{1} & 0 \\ 
0 & \sigma_{2}%
\end{array}
\right ) V^{*}
\end{equation*}
with $U$ and $V$ unitary, and $0\leq \sigma_{2}\leq \sigma_{1}\leq 1$.

This may be rewritten in the form 
\begin{equation*}
A=e^{i(a-b)}S_{1}\left( 
\begin{array}{cc}
\sigma _{1} & 0 \\ 
0 & \sigma _{2}%
\end{array}%
\right) S_{2}^{\ast }
\end{equation*}%
with $S_{i}\in SU\left( 2\right) $.

Since $\bar{B}$ is a square root of $I - A^{*}A$ it must be unitarily
related to the unique positive semidefinite square root $(I-A^{*}A)^{\frac{1%
}{2}}$ of $I - A^{*}A$.

But 
\begin{equation*}
(I-A^{*}A)^{\frac{1}{2}} = V\left (%
\begin{array}{cc}
\theta_{1} & 0 \\ 
0 & \theta_{2}%
\end{array}
\right ) V^{*}
\end{equation*}
where $\theta_{i} = \sqrt{1-\sigma_{i}^{2}}$.

Thus 
\begin{equation*}
B = e^{-ic}\bar{S_{3}}\bar{S_{2}} \left (%
\begin{array}{cc}
\theta_{1} & 0 \\ 
0 & \theta_{2}%
\end{array}
\right )S_{2}^{T}
\end{equation*}

for some $S_{3}\in SU\left( 2\right) $ and some real scalar $c$. Equating $%
A^{\ast }B$ to $B^{T}\bar{A}$ we find 
\begin{equation}
\left( 
\begin{array}{cc}
\sigma _{1} & 0 \\ 
0 & \sigma _{2}%
\end{array}%
\right) S_{1}^{\ast }\bar{S_{3}}\bar{S_{2}}\left( 
\begin{array}{cc}
\theta _{1} & 0 \\ 
0 & \theta _{2}%
\end{array}%
\right) =\left( 
\begin{array}{cc}
\theta _{1} & 0 \\ 
0 & \theta _{2}%
\end{array}%
\right) S_{2}^{\ast }S_{3}^{\ast }\bar{S_{1}}\left( 
\begin{array}{cc}
\sigma _{1} & 0 \\ 
0 & \sigma _{2}%
\end{array}%
\right)   \label{MainEqofAppendix}
\end{equation}

The analysis now is naturally divided into several cases:

\begin{description}
\item[Case 1)] Suppose $\sigma _{1}\sigma _{2}\neq 0,\theta _{1}\theta
_{2}\neq 0$ and $\sigma _{1}\neq \sigma _{2}$:

\item Then, first note $0<\sigma _{2}<\sigma _{1}<1$. Premultiplying both
sides of Equation (\ref{MainEqofAppendix}) by the inverse of $\left( 
\begin{array}{cc}
\theta _{1} & 0 \\ 
0 & \theta _{2}%
\end{array}%
\right) $ to find 
\begin{equation}
DS_{4}^{T}=S_{4}D  \label{FugledePutnamType}
\end{equation}%
where $S_{4}=S_{2}^{\ast }S_{3}^{\ast }\bar{S}_{1}\in SU\left( 2\right) $
and $D=\left( 
\begin{array}{cc}
\gamma _{1} & 0 \\ 
0 & \gamma _{2}%
\end{array}%
\right) $, with $\gamma _{i}=\frac{\sigma _{i}}{\theta _{i}},$ $i=1,$ $2$.

\item Since $S_{4}\in SU\left( 2\right) $ we can write it in so-called
Cayley-Klein form as 
\begin{equation*}
S_{4}=\left( 
\begin{array}{cc}
ce^{i\lambda } & se^{I\mu } \\ 
-se^{-i\mu } & ce^{-i\lambda }%
\end{array}%
\right) 
\end{equation*}%
with $c=\cos (\theta ),$ $s=\sin (\theta )$ for some $\theta \in \lbrack 0,%
\frac{\pi }{2}]$ and $\lambda ,$ $\mu \in \lbrack 0,2\pi ]$. Equation (\ref{%
FugledePutnamType}) now forces $2$ alternatives: $i)$ either $s=0$ or $ii)$ $%
s\neq 0$ and $e^{i2\mu }=-\frac{\gamma _{1}}{\gamma _{2}}$. For the case at
hand, the former alternative holds, since the latter alternative forces $%
\gamma _{1}=\gamma _{2}$ and hence $\sigma _{1}=\sigma _{2}$. So $s=0$ and
hence $S_{4}$ is diagonal $=$ $\left( 
\begin{array}{cc}
e^{i\lambda } & 0 \\ 
0 & e^{-i\lambda }%
\end{array}%
\right) $. So $S_{3}=S_{2}\left( 
\begin{array}{cc}
e^{i\lambda } & 0 \\ 
0 & e^{-i\lambda }%
\end{array}%
\right) S_{1}^{T}$.

\item[Case 2)] $\sigma _{1}\sigma _{2}\neq 0,$ $\theta _{1}\theta _{2}\neq 0$
and $\sigma _{1}=\sigma _{2}$. In this case, as both singular values of $A$
are equal, we have $A=kU$, where $\left\vert k\right\vert <1$ and $U$ is $%
2\times 2$ unitary. Hence $B=\sqrt{1-\left\vert k\right\vert ^{2}}V$ for
some unitary $V$. We still have to impose the requirement that $A^{\ast }B$
is symmetric. To that end, we write $A=e^{ia}S_{1},$ $V=e^{ib}S_{2}$ with $%
S_{j}\in SU\left( 2\right) $, written in Cayley-Klein form as%
\begin{equation*}
S_{J}=%
\begin{pmatrix}
c_{j}e^{i\lambda _{j}} & s_{j}e^{i\mu _{j}} \\ 
-s_{j}e^{-i\mu _{j}} & c_{j}e^{-i\lambda _{j}}%
\end{pmatrix}%
,\text{ }j=1,\text{ }2
\end{equation*}%
with $c_{j}=\cos (\theta _{j}),$ $s_{j}=\sin (\theta _{j}),$ $j=1,$ $2$.
Then $A^{\ast }B$ symmetric is equivalent to 
\begin{equation*}
c_{1}s_{2}\cos (\mu _{2}-\lambda _{1})=s_{1}c_{2}\cos (\lambda _{2}-\mu _{1})
\end{equation*}

\item[Case 3)] $\sigma _{1}=\sigma _{2}=1$. In this case $A$ is unitary and $%
B=0$. Of course, $A^{\ast }B$ is trivially symmetric.

\item[Case 4)] $\sigma _{1}=0$. In this case $A=0$ and $B$ is any unitary
matrix. Once again $A^{\ast }B$ is trivially symmetric.

\item[Case 5)] $\sigma _{2}=0$, but $\sigma _{1}\neq 0$: Now $\theta _{2}=1$%
, while $\theta _{1}\neq 0$. So as $\theta _{1}\theta _{2}\neq 0$ and $%
\gamma _{1}\neq \gamma _{2}$, the analysis for Case 1 still applies to show
that $S_{4}$ is diagonal. Hence, $A=e^{i\phi }S_{1}\left( 
\begin{array}{cc}
\sigma _{1} & 0 \\ 
0 & 0%
\end{array}%
\right) S_{2}^{\ast }$ and $B=S_{1}\left( 
\begin{array}{cc}
\theta _{1}e^{i(\lambda -c)} & 0 \\ 
0 & e^{-i(\lambda +c)}%
\end{array}%
\right) S_{1}^{\ast }$.

\item[Case 6)] Precisely one of the $\theta _{i}=0$: In this case it has to
be $\theta _{1}$, since $\sigma _{1}>\sigma _{2}$. This forces $\sigma
_{1}=1,\sigma _{2}=0$. To analyse this case we rewrite Equation (\ref%
{MainEqofAppendix}) as 
\begin{equation*}
\left( 
\begin{array}{cc}
1 & 0 \\ 
0 & \sigma _{2}%
\end{array}%
\right) S_{4}\left( 
\begin{array}{cc}
0 & 0 \\ 
0 & \theta _{2}%
\end{array}%
\right) =\left( 
\begin{array}{cc}
0 & 0 \\ 
0 & \theta _{2}%
\end{array}%
\right) S_{4}^{T}\left( 
\begin{array}{cc}
1 & 0 \\ 
0 & \sigma _{2}%
\end{array}%
\right) S_{4}
\end{equation*}%
Premultiplying and postmultiplying both sides by the inverse of $\left( 
\begin{array}{cc}
1 & 0 \\ 
0 & \sigma _{2}%
\end{array}%
\right) $ we get $S_{4}D=DS_{4}^{T}$, where $S_{4}=S_{1}^{\ast }\bar{S}_{3}%
\bar{S_{2}}$ and $D=\left( 
\begin{array}{cc}
0 & 0 \\ 
0 & \frac{\theta _{2}}{\sigma _{2}}%
\end{array}%
\right) $. Once again this forces $S_{4}$ to be diagonal. Hence overall $%
A=e^{i(a-b)}S_{1}\left( 
\begin{array}{cc}
1 & 0 \\ 
0 & \sigma _{2}%
\end{array}%
\right) S_{2}^{\ast }$ and $B=e^{-ic}S_{1}\left( 
\begin{array}{cc}
0 & 0 \\ 
0 & \theta _{2}e^{-i\lambda }%
\end{array}%
\right) S_{1}^{\ast }$.
\end{description}

Future work will address the inversion of the covering map in dimensions $5$
and $6$. It is hoped that this characterization of the blocks $A$ and $B$ of
an element of $Sp\left( 4\right) $ leads to a satisfactory solution to the
question of inverting the covering map in dimension $5$, as well as being
useful in other problems in which $Sp(4)$ intervenes.

\end{document}